\documentclass[aps,prb,twocolumn,floats,showpacs]{revtex4}
\usepackage{latexsym}
\usepackage{amsmath}
\usepackage{amssymb}
\usepackage{bm}
\usepackage{wasysym}
\usepackage{color}
\usepackage{graphicx}
\usepackage{subfigure}
\usepackage{flushend}
    \usepackage{lipsum}
   \usepackage{balance}
\usepackage{tikz}

\newcommand{\hide}[1]{}
\newcommand{\veps}{\varepsilon}

\def\veps{\varepsilon}

\def\1S0{$^{1}$S$_{0}$}
\def\3P0{$^3$P$_0$}

\def\g{\mathrm{g}}
\def\e{\mathrm{e}}

\def\mbfr{\mathbf{r}}
\def\mbfR{\mathbf{R}}

\def\mbfn{\mathbf{n}}
\usepackage{amsmath,stmaryrd,graphicx}
\makeatletter
\newcommand{\fixed@sra}{$\vrule height 2\fontdimen22\textfont2 width 0pt\shortrightarrow$}
\newcommand{\shortarrow}[1]{%
  \mathrel{\text{\rotatebox[origin=c]{\numexpr#1*45}{\fixed@sra}}}
}
\makeatother

\begin{document}

\title{Coqblin-Schrieffer Model for an Ultra-cold
       Gas of Ytterbium atoms with Metastable States}

\author{Igor Kuzmenko$^{1}$, Tetyana Kuzmenko$^{1}$,
        Yshai Avishai$^{1,2}$ and Gyu-Boong Jo$^{3}$}
  \affiliation{\footnotesize
  $^{1}$Department of Physics, Ben-Gurion University of
  the Negev, Beer-Sheva, Israel, \\
  $^{2}$NYU-Shanghai, Pudong, Shanghai, China, \\
  $^{3}$Department of Physics, The Hong Kong University
  of Science and Technology, Clear Water Bay, Kowloon,
  Hong Kong, China}

\begin{abstract}
Motivated by the impressive recent advance in
manipulating cold ytterbium atoms we explore
and substantiate the feasibility of realizing
the Coqblin-Schrieffer model in a gas of cold
fermionic $^{173}$Yb atoms. Making use of
different AC polarizabillity of the electronic
ground  state (electronic configuration
$^1$S$_0$) and the long lived metastable
state (electronic configuration $^3$P$_0$),
it is substantiated that the latter can be
localized and serve as a magnetic impurity
while the former remains itinerant. The
exchange mechanism between the itinerant
$^1$S$_0$ and the localized $^3$P$_0$
atoms is analyzed and shown to be antiferromagnetic.
The ensuing SU(6) symmetric Coqblin-Schrieffer
Hamiltonian is constructed, and, using
the calculated exchange constant $J$, perturbative
renormalization group (RG) analysis yields
the Kondo temperature $T_K$ that is experimentally
accessible. A number of thermodynamic measurable
observables are calculated in the weak coupling
regime $T>T_K$ (using perturbative RG analysis)
and in the strong coupling regime $T<T_K$
(employing known Bethe ansatz techniques).
\end{abstract}


\date{\today}

\maketitle

\section{Introduction}
  \label{sec-intro}

Ever since its discovery, the physics exposed in cold
atom systems proves to be a godsend for elucidating
spectacular physical phenomena that are otherwise
extremely difficult to access elsewhere  \cite{CAGen,%
CAFerm,UCA-pra04,UCA-Adv-At-Mol-Opt-Phys06,%
UCA-pra06,UCA-nucl-phys07,UCA-Varena06,%
UCA-nature08,UCA-IntJModPhys09,UCA-Nature10,%
UCA-Science10,ExpExc-nature10,Exp-exch-Science08,%
exch-arxiv-15-07,exch-prl-15,exch-nature-15,Nature03,%
arXiv09,prb09,NaturePhys09,Greiner-PhD-03}.
Special  attention is recently focused on quantum
magnetism in general, and  impurity problems in
particular \cite{KE-UCA-prl-10,KE-UCA-arxiv-15-07,%
KE-UCA-arxiv-15-02,2CKECA,Dem,opt-latt-11,%
Demler-Salomon,ITYK}. One of the reasons is that
a cold atom system opens a way
to study the physical properties of a gas of fermionic
atoms with half-integer spin $s\geq\frac{3}{2}$,
 thereby enabling the study of novel impurity problems.
The main goal of this paper is to
develop this general idea into an experimental and theoretical
framework wherein
the Coqblin-Schrieffer model can be realized in an atomic gas of cold
$^{173}$Yb atoms.

In the ``traditional'' Kondo effect \cite{Kondo-64,%
Coleman-PhysWorld-95,Anderson-PhysWorld-95,%
Hewson-book},  a magnetic
impurity of spin ${\mathbf{S}}$ immersed in a metal
host, scatters the itinerant electrons having spin
${\mathbf{s}}$ ($s=\frac{1}{2}$)
through an antiferromagnetic exchange interaction
$J{\mathbf{s}}\cdot{\mathbf{S}}$ with $J>0$, and
the pertinent dynamics is governed by the $s$-$d$ exchange
Hamiltonian \cite{Hewson-book}.
In the Coqblin-Schrieffer model \cite{Hewson-book,%
Rajan-prl-83,Schlottmann-83,KE-prb-98,Coq-Schr-arxiv-01,%
Coq-Schr-prb-03,Coq-Schr-JPhSocJpn-07,KA-prb14,%
Coq-Schr-Physica14},
the itinerant fermions and the impurity are both $N$-fold
``spin'' degenerate, so that the corresponding Hamiltonian
 has an SU($N$) symmetry. The main difference
between the s-d exchange model for spin $S=\tfrac{1}{2}$ and
the Coqblin-Schrieffer model for ``spin'' $S>\frac{1}{2}$
is that, due to exchange scattering, the change of
the $z$-component of the angular momentum of
the impurity is restricted to $0,\pm1$ in the s-d model,
while it is unrestricted in the Coqblin-Schrieffer
model. In solid state physics, the high level degeneracy
is due to spin-orbit coupling, so that the model is relevant
for applications to rare earth impurities. In a gas of
ultracold atoms, the degeneracy is due solely to
the atomic total angular momentum ${\bf F}={\bf I}+{\bf J}$,
where ${\bf I}$ is the nuclear spin and ${\bf J}$ is the total
(orbital and spin) electronic angular momentum.

Realizing the Coqblin-Schrieffer model in cold
fermionic   $^{173}$Yb atoms is feasible due to a rather unique
exchange mechanism.  The atoms in the $^1$S$_0$
ground-state form a Fermi gas with SU($N$) symmetry and the atoms in
the long-lived $^3$P$_0$  excited state assume
the role of localized magnetic impurities. Both
the ground and excited states have spin
$F=\frac{5}{2}$ (which is the nuclear spin). The idea
is to localize an excited  $^3$P$_0$ atom
in a state-dependent optical potential, such that it
will serve as a magnetic impurity, immersed in a Fermi
gas of ground state $^1$S$_0$ atoms.  The latter is
confined in a combination of harmonic and periodic
potentials but otherwise are itinerant. We show that
an antiferromagnetic exchange interaction exists
between the itinerant and localized atoms and that
the ensuing exchange scattering is described by
the Coqblin-Schrieffer Hamiltonian.

In Sec. \ref{Yt} we briefly review the advantage of using
degenerate alkaline-earth-like atoms such as Yb and Sr
for the study the Kondo effect and its SU($N$) extension
in cold atom experiments. Then, in Sec. \ref{sec-description},
we present a general description of the system composed
of a mixture of $^{173}$Yb atoms in their ground and
excited states. Atomic quantum states in the optical
potential are described in Subsec. \ref{sec-model},
while the exchange interaction between Yb atoms in
the ground and excited states is derived in
Subsec. \ref{subsec-exchange}. This exchange
interaction is somewhat unusual because it occurs
between the same atoms whose electronic angular
momentum is zero. In Subsec. \ref{subsec-HK} we
derive the SU($N$) Kondo Hamiltonian and estimate
the Kondo temperature. Calculations of observables
are detailed in Sec. \ref{sec-observables}, and
naturally divided into the weak and strong coupling
regimes. The magnetic susceptibility, entropy and
specific heat of the impurity, in the weak coupling
regime (${T}>{T}_{K}$) are estimated in Subsection
\ref{sec-suscept-weak}. Magnetic susceptibility,
entropy and specific heat of the impurity in the strong
coupling regime ($T<T_K$) are derived in
Subsecsection \ref{sec-strong}. Our main results
are summarized in Sec. \ref{sec-conclusions}.
Details of the derivation of the exchange interaction
between two Yb atoms in $^1$S$_0$ and $^3$P$_0$
respective atomic states are expanded upon in
the Appendix. It is shown and underlined there
that precise calculation of the exchange constant
requires a detailed knowledge of the atomic wave
functions. Although these details are of technical nature,
they expose how the exchange interaction determines
the scattering length, and demonstrate the extreme
sensitivity of the relation between the singlet and
triplet scattering lengths on the one hand and
the magnitude of the exchange interaction on
the other hand.
Indirect exchange
interaction is considered in subsection
\ref{append-hybridization}. It is shown that
the exchange is antiferromagnetic. This conclusion
does not depend on a chosen model or
an approximation but is general property of
the second order perturbation theory. Wave function
describing motion of interacting atoms is derived in
subsection \ref{subsection-WKB}. We derive here
expression for the scattering length. In subsection
\ref{subsec-comparison} we compare our results for
the scattering length with experimental results of
Ref. \cite{6}.
In subsection \ref{subsec:CIR} we study
confinement-induced resonances (CIR) where
the exchange interaction changes from ferromagnetic
to antiferromagnetic and vise versa.

\section{Kondo effect with cold alkaline atoms}
  \label{Yt}

Recent advance in the techniques of cooling and
manipulating degenerate alkaline-earth-like
atoms (e.g. ytterbium and/or strontium atoms) \cite{1,2,3}
paves the way for studying  novel aspects of
interacting Fermi systems.  These include
non-equilibrium properties such as transport, as well
as impurity problems and other facets of quantum
magnetism. A key role in these considerations is
played by the interplay between the long-lived
metastable $^3$P$_0$ state and the $^1$S$_0$
ground state,  with their enlarged SU($N$) spin
symmetry for fermionic isotopes~\cite{add1,add2}.
Utilizing a narrow singlet-triplet optical transition,
for example, alkaline-earth-like atoms have been
thought of as a promising candidate for realizing
a precise atomic clock~\cite{add3} or ideal
storage of qubits for the application of quantum
computing~\cite{add2}. 

Here we consider the possible
occurrence of the Kondo effect and its SU($N$) extensions
in a gas of ytterbium atoms.  Making use of
different AC polarizabilities of the ground-state ($^1$S$_0$)
and the long lived metastable ($^3$P$_0$) state,
the localized $^3$P$_0$ atoms can serve as magnetic
impurities which interact with itinerant ground-state
atoms~\cite{4,4-1,4-2}. The Kondo effect arises when
this interaction is characterized by spin-exchange
between $^1$S$_0$ and  $^3$P$_0$  state. Such
spin-exchange interactions has  recently been
demonstrated\cite{5,6,7} in fermionic $^{173}$Yb atoms.

Realizing the Kondo effect in alkali-earth-like atoms
exposes novel aspects of the Kondo physics with SU($N$)
symmetric interactions that are difficult to elucidate in
solid-state based system,  because the high SU($N$)
symmetry arises from the strong decoupling between
nuclear and electronic spins in alkali-earth-like atoms.
As such, it has attracted much interest in the context of
SU($N$) Fermi gases both for bulk systems~\cite{8,9,10}
and for lattice systems~\cite{11}. Here, we focus on
the SU($N$) Kondo model in the fermionic $^{173}$Yb
gas, and estimate the Kondo temperature. In cold atom
systems, due to the weak magnetic coupling of
spin-exchange interactions, the questions still
remains whether or not the Kondo temperature is
attainable by current experiments. Our finding shows
that  the Kondo temperature is enhanced by the SU($N$)
interactions. In electronic systems, this is shown in
previous works on heavy fermion systems
\cite{Hewson-book,Coleman} and on carbon nanotube
quantum dots \cite{KA-prb14}. Indeed, the Kondo
temperature in cold atom system may also be enhanced
by means of the confinement-induced resonance~\cite{11}
or by the orbital-induced Feshbach resonance
\cite{12,13,14,SU3-KE-cold-atoms-PRL13}.

\section{Description of the System}
  \label{sec-description}

\begin{figure}[htb]
\centering
\includegraphics[width=65 mm,angle=0]
   {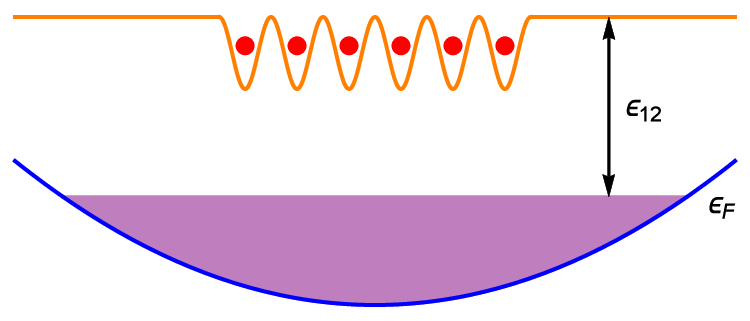}
 \caption{Illustration of a ``Kondo model''
   for  $^{173}$Yb atoms. Atoms in the ground-state
   $^1$S$_0$ form a Fermi sea, while atoms  in
   the excited-state $^3$P$_0$ are trapped in an optical
   potential and form a dilute concentration of localized
   magnetic impurities (see details in the text).}
 \label{Fig-model}
\end{figure}

Having underlined the peculiar advantage of using alkaline
atoms to explore the Kondo effect and its SU($N$) extensions,
we now focus a cold gas of $^{173}$Yb fermionic
atoms confined in a shallow harmonic potential. Most of
the atoms remain in the ground state $^{1}$S$_{0}$ and
form a Fermi sea due to its half integer nuclear spin $I=\tfrac{5}{2}$
(purple area in Fig. \ref{Fig-model}).  However, a few
atoms are found in a long lived excited $^{3}$P$_{0}$ state following a
coherent excitation via the clock transition. These excited
atoms can be trapped in a state-dependent optical lattice
potential as schematically displayed  in Fig. \ref{Fig-model}
(red circles), and can
be regarded as localized impurities. The wavelength
of the periodic optical potential exceeds the range
of interaction between atoms, thereby justifying
the assumption that the concentration of excited atoms
is small enough so that they are not correlated.

\vspace{0.1in}

In the following, we describe such mixture of $^{173}$Yb
atomic system within a model of uncorrelated and localized
magnetic impurities. To this end,  the details of an exchange
interaction between an atom in the ground-state and
an atom in an excited state is of crucial importance.
Since both atoms in the ground and excited states
are in an electronic singlet state, direct exchange
interaction between these atoms is absent. There is,
however, an indirect exchange,  that involves virtual
hopping of electrons between the atoms such that
an atom transforms from the ground state to an excited
state, whereas the other atom transforms from an excited
state to the ground state. An expression for this exchange
interaction is derived below, followed by an analysis of
the corresponding impurity problem, that turns out to be
a manifestation of the Coqblin-Schrieffer model realized
in cold atom systems.

\subsection{Quantum States of $^{173}$Yb Atoms
  with van der Waals Interaction}
  \label{sec-model}

\begin{figure}[htb]
\centering
\includegraphics[width=55 mm,angle=0]
   {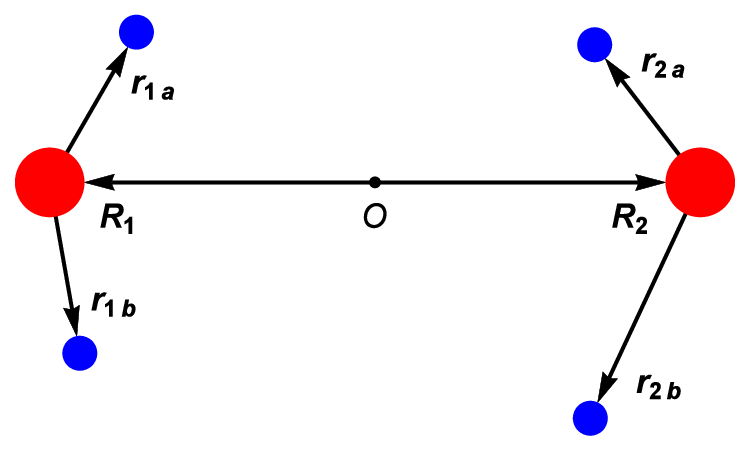}
 \caption{\footnotesize
   Two ytterbium atoms.
   The position of the rigid ions are $\mbfR_1$
   and $\mbfR_2$,
   positions of the electrons are ${\bf{r}}_{1{\mathrm{a}}}$,
   ${\bf{r}}_{1{\mathrm{b}}}$, ${\bf{r}}_{2{\mathrm{a}}}$ and
   ${\bf{r}}_{2{\mathrm{b}}}$. The origin of the frame is
   denoted as $O$.}
   \vspace{-0.1in}
 \label{Fig-atoms}
\end{figure}

Before discussing exchange interaction between two Yb atoms 
it is important to analize the single atom properties because the 
exchange interaction is crucially dependent on the electronic wave functions 
of a single atom. 
An $^{173}$Yb atom can be considered as a charged (+2)
closed shell rigid ion and two valence electrons.
The ground-state $^{1}$S$_{0}$ valence electrons
configuration is $6s^2$, while that of the excited
state $^{3}$P$_{0}$ is 6$s$6$p$. The excitation
energy $\epsilon_{12}=\epsilon_2-\epsilon_1$
is \cite{spectral-line-Yb}
\begin{eqnarray*}
  \epsilon_{12}
  &=&
  2.14349~
  {\mathrm{eV}}.
\end{eqnarray*}
The positions of the ion core and the outer electrons are
respectively specified by vectors $\mbfR$,
$\mbfr_{\mathrm{a}}$ and $\mbfr_{\mathrm{b}}$
(Fig. \ref{Fig-atoms}).

The ytterbium atoms are trapped by state-dependent
trapping potentials $V_{\g,\e}(\mbfR)$,
\begin{eqnarray}
  V_{\g}(\mbfR) &=&
  V_{\g}^{(0)}
  k_{\g}^{2}
  R^2,
  \ \ \
  R~=~|\mbfR|,
  \label{V-g-trap}
  \\
  V_{\e}(\mbfR) &=&
  V_{\e}^{(0)}
  \sum_{i}
  \sin^2
  \big(
      k_{\e}
      X_i
  \big),
  \label{V-x-trap}
\end{eqnarray}
where $i$ is a Cartesian index.
The potential parameters are tuned such that
\begin{eqnarray}
  V_{\g}^{(0)}
  k_{\g}^2
  ~\ll~
  V_{\e}^{(0)}
  k_{\e}^{2},
  \label{inequalities}
\end{eqnarray}
and therefore the atoms in the ground state are considered
as itinerant atoms, and the atom in the excited state plays
a role of the impurity.

In the adiabatic (Born-Oppenheimer) approximation
(which is well substantiated in atomic physics),
the wave function of a single ytterbium atom is
expressed as a product of the wave functions
$\Psi(\mbfR)$ (for the rigid ion core) and
$\psi(\mbfr_a,\mbfr_b)$ (for the valence electrons).
The former is considered as a point particle of mass
$M$ whose position vector in Cartesian coordinates
is ${\bf R}=(X,Y,Z)$.

Starting with the core wave functions, recall that
the atoms in the ground-state and the excited
state are subject to different 3D optical potentials
and van der Waals interactions between the atoms.
Strictly speaking, we should describe the system
by many-particle wave function
$\Psi(\mbfR_0;\{\mbfR\}_{\cal{N}})$, where ${\cal{N}}$
is the number of itinerant atoms,
$\{\mbfR\}_{\cal{N}}=\{\mbfR_1,\mbfR_2,\ldots,\mbfR_{\cal{N}}\}$,
$\mbfR_j$ is the position of an itinerant atom
($j=1,2,\ldots,{\cal{N}}$) and
$\mbfR_0$ is the position of the impurity atom. When the distance
between all the atoms exceeds the range of the vad der Waals
interaction, the many-particle wave function splits into a product
of single-particle wave functions. When an itinerant atom is placed
close to the impurity and all other atoms are far away, the many
particle wave function is a product of a two-particle wave
function describing interacting pair of atoms, and single
particle wave functions describing motion of the other itinerant
atoms. Usually, the density of itinerant atoms is low and
the probability to find two or more itinerant atoms close to
the impurity is negligible small. Therefore, we can describe
the many atomic system in terms of two-atomic wave functions.
For this purpose we use the notations $\Psi(\mbfR_1,\mbfR_2)$
for the core wave functions pertaining for two
atoms in the ground or excited electronic states.
They are solutions of the following Schr\"odinger
equation:
\begin{eqnarray}
  {\cal{H}}
  \Psi(\mbfR_1,\mbfR_2)
  &=&
  {\cal{E}}
  \Psi(\mbfR_1,\mbfR_2).
  \label{eq-Schrodinger-2atoms}
\end{eqnarray}
Here $\mbfR_1$ is the position of the atom in the ground
state, $\mbfR_2$ is the position of the atom in the excited
state. The two particle Hamiltonian ${\cal{H}}$ is,
\begin{eqnarray}
  {\cal{H}} &=&
  {\cal{H}}_{\g}+
  {\cal{H}}_{\e}+
  W\big(|\mbfR_1-\mbfR_2|\big).
  \label{H-2atoms}
\end{eqnarray}
The first or second terms on the right hand side of
eq. (\ref{H-2atoms}) describe motion of the atom in
the groung or excited state,
\begin{eqnarray}
  {\cal{H}}_{\g} &=&
  -\frac{\hbar^2}{2M}~
  \frac{\partial^2}{\partial\mbfR_1^2}+
  V_{\g}(\mbfR_1),
  \label{H-GS-def}
  \\
  {\cal{H}}_{\e} &=&
  -\frac{\hbar^2}{2M}~
  \frac{\partial^2}{\partial\mbfR_2^2}+
  V_{\e}(\mbfR_2),
  \label{H-ES-def}
\end{eqnarray}
where $M$ is the atomic mass. Recall that the trapping
potentials $V_{\g,\e}(\mbfR)$ are defined in Eqs.
(\ref{V-g-trap}) and (\ref{V-x-trap}).

The third term on the right hand side of eq. (\ref{H-2atoms})
is the Van der Waals interaction between the ytterbium atoms.
Explicitly, it is expressed as, \cite{vdW-Yb-PRA-08,vdW-Yb-PRA-14}
\begin{eqnarray}
  W(R) &=&
  \frac{C_6}{R^6}~
  \bigg\{
       \frac{\sigma^6}{R^6}-1
  \bigg\}-
  \frac{C_8}{R^8}.
  \label{vdW-def}
\end{eqnarray}
Here $C_6=2.651\cdot10^3E_ha_B^6$,
$C_8=3.23640441\cdot10^5E_ha_B^8$ and
$\sigma=9.0109361a_B$, where
$E_h=27.211$~eV is the Hartree energy and
$a_B=0.52918$~{\AA} is the Bohr radius.

\begin{figure}[htb]
\centering
\includegraphics[width=65 mm,angle=0]
   {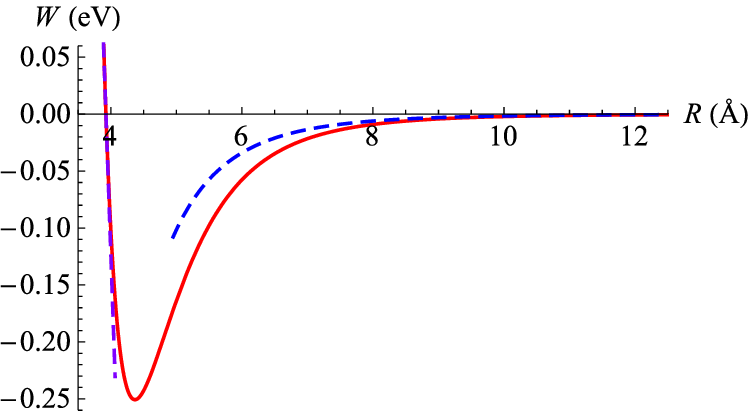}
 \caption{Van der Waals interatomic interaction (solid red curve),
   its approximation by $-C_6/R^6$ (blue dashed curve)
   and linear approximation near the classical turning point
   $R=r_0$ (dashed purple line).}
 \label{Fig-vdW}
\end{figure}

Van der Waals potential is illustrated in Fig. \ref{Fig-vdW}. It
is equal to zero when $R=r_0$, where $r_0=3.9$~{\AA}. At
$R = a_0 = 4.4$~{\AA}, the potential $W(R)$ reaches its minimum,
$W(a_0)=-0.25$~eV.

\subsubsection{Wave Function at Long Distances between
         the Atoms}

There is a characteristic length $\lambda$ associated with
the van der Waals potential,
\begin{eqnarray}
  \lambda &=&
  \bigg(
       \frac{M C_6}{\hbar^2}
  \bigg)^{1/4}
  ~=~
  89.97~{\text{\AA}}.
  \label{lambda-def}
\end{eqnarray}
When the distance between the atoms exceeds $\lambda$,
we can neglect the van der Waals interaction. In this case
the two atomic wave function $\Psi(\mbfR_1,\mbfR_2)$
takes the form of a product of two single atomic wave
functions, $\Psi_{\g}(\mbfR_1)$ and $\Psi_{\e}(\mbfR_2)$,
\begin{eqnarray*}
  \Psi(\mbfR_1,\mbfR_2) &=&
  \Psi_{\g}(\mbfR_1)
  \Psi_{\e}(\mbfR_2).
\end{eqnarray*}
The wave functions $\Psi_{\g,\e}(\mbfR)$ are
respectively the eigenfunctions of the Hamiltonians
${\cal{H}}_{\g}$ and ${\cal{H}}_{\e}$,
that is, 
\begin{equation}
  -\frac{\hbar^2}{2M}~
  \frac{\partial^2 \Psi_{\g,\e}(\mbfR)}
       {\partial \mbfR^2}+
  V_{\g,\e}(\mbfR)
  \Psi_{\g,\e}(\mbfR)=
  \veps
  \Psi_{\g,\e}(\mbfR).
  \label{eq-Schrodinger-atom}
\end{equation}
Consider first the wave function $\Psi_{\e}(\mbfR)$.
When the corresponding energy level
$\veps_{\mathrm{imp}}$ is deep enough, the wave
function of the bound state near the potential
minimum at $\mbfR=0$ can be approximated within
the harmonic potential picture as
\begin{eqnarray}
  \Psi_{\e}(\mbfR) &=&
  \frac{1}{\big(\pi b_{\e}^2\big)^{3/4}}~
  \exp
  \bigg(
       -\frac{R^2}{2 b_{\e}^{2}}
  \bigg),
  \label{WF-atom-x}
\end{eqnarray}
where the harmonic length and frequency are
\begin{eqnarray}
  &&
  k_{\e}
  b_{\e}
  =
  \bigg[\frac{\hbar^2 k_{\e}^{2}}{2 M V_{\e}^{(0)}}\bigg]^{1/4},
  \ \
  \Omega_{\e}
  =
  \sqrt{\frac{2 V_{\e}^{(0)} k_{\e}^{2}}{M}}.
  \label{a-x-omega-x}
\end{eqnarray}
Energy level of the impurity is
\begin{eqnarray}
  \veps_{\mathrm{imp}}
  =
  \frac{3\hbar\Omega_{\e}}{2}.
  \label{epsilon-x}
\end{eqnarray}
Next, consider the wave functions $\Psi_{\g}(\mbfR)$
of the ytterbium atom in the ground state for which the shallow
potential wells are not deep enough to form tightly bound states.
Hence, we can neglect the ``fast'' potential relief and take into
account just isotropic potential $V_{\g}(\mbfR)$.
Quantum states of atoms in isotropic potential (\ref{V-g-trap})
are described by the radial quantum number $n$ [$n=0,1,2,\ldots$],
orbital quantum number $L$ [$L=0,1,2,\ldots$] and projection $m$
of the orbital moment on the axis $z$ [$m=-L,-L+1,\ldots,L$].
Due to the centrifugal barrier, only the atoms with $L=0$ can
approach the impurity and be involved in the exchange interaction
with it. The wave functions of the states with $L=0$ found from
the Schr\"odinger equation (\ref{eq-Schrodinger-atom}) are,
\begin{eqnarray}
  \Psi_{\g}^{n}(\mbfR)
  =
  {\cal{N}}_{n}~
  L_{n}^{(\frac{1}{2})}
  \bigg(
       \frac{R^2}{b_{\g}^{2}}
  \bigg)
  \exp
  \bigg(
       -\frac{R^2}{2 b_{\g}^{2}}
  \bigg),
  \label{WF-g-3D}
\end{eqnarray}
where $L_{n}^{(l+\frac{1}{2})}(\varrho)$ are generalized
Laguerre polynomials. The normalization factor is:
$$
  {\cal{N}}_{n} ~=~
  \bigg(
       \frac{1}{2 \pi b_{\g}^{2}}
  \bigg)^{3/4}~
  \sqrt{\frac{2^{n+2}~n!}{(2n+1)!!}}.
$$
The harmonic length $b_{\g}$ and frequency $\Omega_{\g}$
are defined through,
\begin{eqnarray}
  k_{\g} b_{\g}
  ~=~
  \bigg(\frac{\hbar^2 k_{\g}^{2}}{2 M V_{\g}^{(0)}}\bigg)^{1/4},
  \ \ \
  \Omega_{\g} ~=~
  \sqrt{\frac{2 V_{\g}^{(0)} k_{\g}^{2}}{M}}.
  \label{a-g-omega-g-def}
\end{eqnarray}
The corresponding energy levels are
\begin{eqnarray}
  \veps_{n} &=&
  \hbar \Omega_{\g}
  \bigg(
       2n+
       \frac{3}{2}
  \bigg).
  \label{energy-levels-atom}
\end{eqnarray}
The inequality (\ref{inequalities}) imply
$$
  \Omega_{\g}
  ~\ll~
  \Omega_{\e}.
$$
Within this framework, the spectrum is nearly continuous
and the ytterbium atoms in the ground-state form a Fermi
gas. The Fermi energy $\epsilon_F$ is such that
$\epsilon_F\gg\hbar\Omega_{\g}$, hence the Fermi gas is 3D.

\subsubsection{Wave Function at Short Distances between
               the Atoms}

In order to elucidate the behavior of the two atomic wave function
within the interval $|\mbfR_1-\mbfR_2|\lesssim\lambda$, we
adopt the semiclassical technique developed in Ref.
\cite{Scatt-Length-WKB-PRA93}: Introduce the coordinate
$\mbfR_c$ of the center of mass and the relative coordinate
$\mbfR$,
\begin{eqnarray}
  \mbfR_c ~=~
  \frac{1}{2}~
  \big(
      \mbfR_1+
      \mbfR_2
  \big),
  \ \ \ \ \
  \mbfR ~=~
  \mbfR_1-
  \mbfR_2.
  \label{R-com-R-relative}
\end{eqnarray}
In the next step, we take into account that the van der Waals
potential (\ref{vdW-def}) has an effective range $\lambda$,
Eq. (\ref{lambda-def}), and therefore $|\mbfR| \lesssim \lambda$.
Employing the following inequalities,
\begin{eqnarray*}
  k_{\g} \lambda
  ~\ll~
  k_{\e}\lambda
  ~\ll~ 1,
\end{eqnarray*}
we can write
\begin{eqnarray*}
  V_{\e}
  \bigg(
       \mbfR_c -
       \frac{\mbfR}{2}
  \bigg)
  &=&
  V_{\e}\big(\mbfR_c\big)+
  O\big(k_{\e}\lambda\big),
  \\
  V_{\g}
  \bigg(
       \mbfR_c +
       \frac{\mbfR}{2}
  \bigg)
  &=&
  V_{\g}^{(0)}~
  k_{\g}^{2} R_{c}^{2}.
\end{eqnarray*}
The motion of the atom in the excited state is restricted
within the area $|\mbfR_2| \approx |\mbfR_c| \lesssim b_{\e}$.
Taking into account the inequality $k_{\g} b_{\e} \ll 1$,
we can neglect $V_{\g}(R)$ within the intervals,
$$
  \big|\mbfR_2\big|
  ~\lesssim~
  b_{\e},
  \ \ \ \ \
  \big|
      \mbfR_1-
      \mbfR_2
  \big|
  ~\lesssim~
  \lambda.
$$
Then the two atomic wave function is a product of two
functions, $\Psi_{c}(\mbfR_c)$ and $\Psi_{r}(\mbfR)$,
which satisfy the equations,
\begin{eqnarray}
  \bigg\{
       -\frac{\hbar^2}{4M}~
       \frac{\partial^2}{\partial\mbfR_c^2}+
       V_{\e}\big(\mbfR_c\big)
  \bigg\}
  \Psi_{c}(\mbfR_c)
  =
  {\cal{E}}_{c}
  \Psi_{c}(\mbfR_c),
  \label{Schrodinger-eq-COM}
  \\
  \bigg\{
       -\frac{\hbar^2}{M}~
       \frac{\partial^2}{\partial\mbfR^2}+
       W(R)
  \bigg\}
  \Psi_{r}(\mbfR)
  =
  {\cal{E}}_{r}~
  \Psi_{r}(\mbfR).
  ~~~~~
  \label{Schrodinger-eq-relative}
\end{eqnarray}
Effect of the trapping potential $V_{\e}(\mbfR_2)$ on the relative
motion of the atoms is considered in Appendix \ref{subsec:CIR}.
Eq. (\ref{Schrodinger-eq-COM}) yields the wave function of a
bound state near the minimum of
$V_{\e}(\mbfR)$ at  $\mbfR=0$.
Before analyzing the wave-function  $\Psi_{r}(\mbfR)$, we note
that the total energy of the two atom system is
$$
  {\cal{E}}_{n} ~=~
  {\cal{E}}_{c} +
  {\cal{E}}_{r}.
$$
On the other side, this same quantity is also given as:
$$
  {\cal{E}}_{n} ~=~
  \veps_{\mathrm{imp}}+
  \veps_{n},
$$
where $\veps_{n}$ is given by eq. (\ref{energy-levels-atom}).
For the degenerate Fermi gas, $\veps_{n} \leq \epsilon_F$ and
$\hbar \Omega_{\e} < \epsilon_F$.
Usually the Fermi energy $\epsilon_F$ is such that the Fermi
temperature $T_F=\epsilon_F/k_B$ lies within the interval
[see Ref. \cite{5}, for example]
$$
  T_F ~<~ 300~{\text{nK}}.
$$
The depth of the van der Waals potential is
$W(a_0)=0.25$~{eV} [see Fig. \ref{Fig-vdW}].
Then we can neglect ${\cal{E}}_{r}\sim\veps_{n}$
with respect to the van der Waals potential
(\ref{vdW-def}) at the distances $R\lesssim\lambda$.
Then the Schr\"odinger equation
(\ref{Schrodinger-eq-relative}) takes the form,
\begin{eqnarray}
  &&
  \bigg\{
       -\frac{\hbar^2}{M}~
       \frac{\partial^2}{\partial\mbfR^2} +
       W(R)
  \bigg\}~
  \Psi_{r}
  \big(
      \mbfR
  \big)
  ~=~ 0.
  \label{eq-Schrodinger-2atoms-small-R}
\end{eqnarray}

The potential $W(R)$ depends just on the distance $R$
from the impurity. Therefore, the orbital momentum $L$
and its projection $m$ on the axis $z$ are good quantum
numbers. Because of the centrifugal barrier, just atoms with
$L=0$ can approach close one to another.
Therefore we restrict ourselves by considering just
the s-wave (i.e., the wave with $L=0$).
Solution of the equation (\ref{eq-Schrodinger-2atoms-small-R})
is evident but rather cumbersome [see Ref. 
\cite{Scatt-Length-WKB-PRA93} and subsection
\ref{subsection-WKB} for details]. The wave function of
the s-wave satisfying eq. (\ref{eq-Schrodinger-2atoms-small-R})
is
\begin{eqnarray}
  \Psi_{n}(\mbfR) &=&
  \frac{\psi_{n}(R)}{\sqrt{4\pi}~R},
  \label{WF-partial}
\end{eqnarray}
where $n$ is the harmonic quantum number defined by
eq. (\ref{energy-levels-atom}).
In order to find the radial wave function $\psi_{n}(R)$,
it is useful to employ different approximations 
in several corresponding intervals as defined below. 
To this end, we underline the following
constraints on the parameters $R$: $r_0$, $b_0$ and
$\lambda$ as follows: 
\begin{itemize}
\item $r_0$ is determined from the equation $W(r_0)=0$.
      The classical mechanics allows motion of the zero-energy
      particle in the interval $R>r_0$.

\item $b_0$ is constrained by the inequality,
      $$
        \bigg|
             \frac{\sigma^6}{b_0^6}-
             \frac{C_8}{C_6 b_0^2}
        \bigg|
        ~\ll~ 1.
      $$
      For ${R}\geq{b}_{0}$, we can approximate
      $W(R)\approx-C_6/R^6$. Practically, we take $b_0\approx10$~{\AA}
      [see Fig. \ref{Fig-vdW}].

\item $\lambda=(MC_6/\hbar^2)^{1/4}=89.97$~{\AA}. In principle, 
 the Wentzel-Kramers-Brillouin (WKB) approximation
      can be used for for $R \ll \lambda$. 
\end{itemize}
A brief list of approximations per intervals 
is as follows (see details below): 
For the interval $r_0 < R \ll \lambda$, we can apply the WKB
approximation to solve the Schr\"odinger equation
(\ref{Schrodinger-eq-relative}). For the interval
$b_0 < R < \lambda$, we can approximate $W(R)$ by
$-C_6/R^6$ and solve eq. (\ref{Schrodinger-eq-relative}).
The interval $R < r_0$ corresponds to classically
forbidden region where the wave function decays exponentially. 
In the following discussions, we find the wave
function within each interval.
 The intervals $r_0<R\ll\lambda$ and
$b_0<R<\lambda$ overlap one with another since
there is a wide interval $b_0<R\ll\lambda$ where
both the WKB approximation and the approximation
$W(R)\approx-C_6/R^6$ are valid. Therefore, within
this interval both the approaches should give
the same solution. We use this condition as as
a connection condition for the solutions within
two overlapping intervals.

\noindent
{\textbf{1}}. {\underline{Interval $r_0<R\ll\lambda$}}:
The wave function calculated within the WKB approximation
with quantum corrections
\cite{Scatt-Length-WKB-PRA93,vdW-Yb-PRA-08} is,
\begin{eqnarray}
  \psi_{n}^{(1)}(R) &=&
  \frac{A_{1n}}{\sqrt{K(R)}}~
  \sin
  \Big(
      \Phi_{r}(R)+
      \frac{\pi}{4}
  \Big).
  \label{psi-R0<R<b0}
\end{eqnarray}
where
\begin{eqnarray}
  \Phi_{r}(R) &=&
  \int\limits_{r_0}^{R}
  K(R')dR',
  \label{Phi-def}
  \\
  K(R) &=&
  \frac{1}{\hbar}~
  \sqrt{-M W(R)}.
  \label{K-def}
\end{eqnarray}

When the distance between the atoms exceeds $\lambda$,
the interaction between the atoms can be neglected and
the two-atomic wave function is a product of the single-atomic
wave functions (\ref{WF-atom-x}) and (\ref{WF-g-3D}).
The wave function $\psi_n(R)$ and its
derivative $\psi'_{n}(R)$ are continues at $R=\lambda$.
These conditions give
\begin{eqnarray}
  A_{1n} &=&
  \frac{2~
        \sqrt{k_n \lambda}}
       {\pi  a_{\g}}~
  \Gamma\bigg(\frac{3}{4}\bigg)~
  \sqrt{1+\Big(\frac{a_w-\bar{a}}{\bar{a}}\Big)^{2}},
  \label{A1-res}
\end{eqnarray}
where
\begin{eqnarray}
  k_n ~=~
  \frac{2\sqrt{n}}{a_{\g}},
  \label{kn-def}
\end{eqnarray}
the parameters $a_w$ and $\bar{a}$ are given by
Eqs. (\ref{scattering-length}) and (\ref{bar-a}) below.

\noindent
{\textbf{2}}. {\underline{Interval $R>b_0$}}:
Within this interval, we can approximate the potential energy
by $W(R)\approx-C_6/R^6$. The wave function $\psi_{n}^{(2)}(R)$
fir this interval is,
\begin{eqnarray}
  \psi_{n}^{(2)}(R) &=&
  A_{2n}~
  \tilde\psi_{2A}(R)+
  B_{2n}~
  \tilde\psi_{2B}(R),
  \label{psi-b0<R<rk}
\end{eqnarray}
where $\tilde\psi_{2A}(R)$ and $\tilde\psi_{2B}(R)$ are,
\begin{subequations}
\begin{eqnarray}
  \tilde\psi_{2A}(R) &=&
  \sqrt{\frac{2R}{\lambda}}~
  J_{1/4}
  \bigg(
       \frac{\lambda^2}{2R^2}
  \bigg),
  \label{psi-A}
  \\
  \tilde\psi_{2B}(R) &=&
  \sqrt{\frac{2R}{\lambda}}~
  J_{-1/4}
  \bigg(
       \frac{\lambda^2}{2R^2}
  \bigg).
  \label{psi-B}
\end{eqnarray}
  \label{subeqs-psi-AB}
\end{subequations}

There is a large interval $b_0<R\ll\lambda$, where we can
approximate $W(R)$ by $-C_6/R^6$ and apply the WKB
approximation \cite{Scatt-Length-WKB-PRA93}. Therefore,
we can apply the following connection conditions:
For any $R$ within the interval $b_0<R\ll\lambda$,
the equality $\psi_{n}^{(1)}(R)=\psi_{n}^{(2)}(R)$ is valid.
This conditions gives,
\begin{subequations}
\begin{eqnarray}
  A_{2n}
  &=&
  -A_{1n}~
  \frac{\sqrt{\pi \lambda}}{2}~
  \cos
  \Big(
      \Phi_w+
      \frac{\pi}{8}
  \Big),
  \label{A2-vs-A1}
  \\
  B_{2n}
  &=&
  A_{1n}~
  \frac{\sqrt{\pi \lambda}}{2}~
  \sin
  \Big(
      \Phi_w+
      \frac{3\pi}{8}
  \Big),
  \label{B2-vs-A1}
\end{eqnarray}
  \label{subeqs-A2-B2-vs-A1}
\end{subequations}
where
\begin{eqnarray}
  \Phi_w &=&
  \int\limits_{r_0}^{\infty}
  K(R)~dR.
  \label{Phi-w}
\end{eqnarray}

\begin{figure}[htb]
\centering
\includegraphics[width=65 mm,angle=0]
   {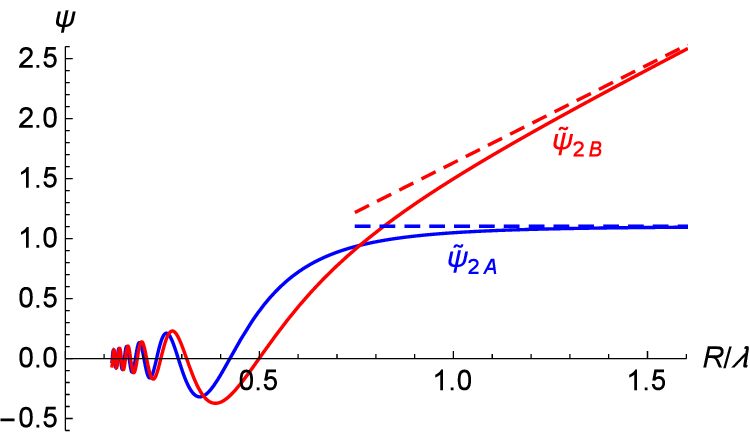}
 \caption{Wave functions $\tilde\psi_{2A}(R)$ and $\tilde\psi_{2B}(R)$,
   eq. (\ref{subeqs-psi-AB}) [solid curves] and
   their asymptotic (\ref{subeqs-psi-AB-R-large}) [dashed lines].}
 \label{Fig-WF-Bessel}
\end{figure}

The functions $\tilde\psi_{2A}(R)$ and $\tilde\psi_{2B}(R)$,
eq. (\ref{subeqs-psi-AB}) are shown in Fig. \ref{Fig-WF-Bessel},
solid lines. It is seen that for $R>\lambda$, the functions
$\tilde\psi_{2A}(R)$ and $\tilde\psi_{2B}(R)$ are well
approximated by the linear with $R$ expressions shown
by dashed lined. Explicitly, the asymptotic expressions for
$\tilde\psi_{2A}(R)$ and $\tilde\psi_{2B}(R)$ are,
\begin{subequations}
\begin{eqnarray}
  \tilde\psi_{2A}(R) =
  \frac{1}{\Gamma(5/4)}~
  \Bigg\{
       1-
       \frac{\lambda^4}{20R^4}+
       O\bigg(\frac{\lambda^8}{R^8}\bigg)
  \Bigg\},
  \label{psi-A-R-large}
  \\
  \tilde\psi_{2B}(R) =
  \frac{2R/\lambda}{\Gamma(3/4)}~
  \Bigg\{
       1-
       \frac{\lambda^4}{12R^4}+
       O\bigg(\frac{\lambda^8}{R^8}\bigg)
  \Bigg\}.
  \label{psi-B-R-large}
\end{eqnarray}
  \label{subeqs-psi-AB-R-large}
\end{subequations}

Eqs. (\ref{psi-b0<R<rk}) and (\ref{subeqs-psi-AB-R-large})
show that the asymptote of the wave function
$\psi_{n}^{(2)}(R)$ at $R\gtrsim\lambda$ is
$\psi_{n}^{(2)}(R)\propto{R-a_w}$, with the scattering
length $a_w$ being \cite{Scatt-Length-WKB-PRA93,vdW-Yb-PRA-08},
\begin{eqnarray}
  a_w &=&
  \bar{a}~
  \bigg\{
       1-
       \tan
       \Big(
           \Phi_w-
           \frac{\pi}{8}
       \Big)
  \bigg\},
  \label{scattering-length}
\end{eqnarray}
where
\begin{eqnarray}
  \bar{a} &=&
  \frac{\lambda}{2^{3/2}}~
  \frac{\Gamma\big(\frac{3}{4}\big)}
       {\Gamma\big(\frac{5}{4}\big)}
  ~=~
  43.01~{\text{\AA}}.
  \label{bar-a}
\end{eqnarray}

For the potential (\ref{vdW-def}), $\Phi_w=294.273$ and
\begin{eqnarray}
  a_w &=&
  650.6~{\text{\AA}}.
  \label{a-vdW-res}
\end{eqnarray}

\subsection{Exchange Interaction}
  \label{subsec-exchange}

The strength of the exchange interaction between Yb(\1S0) and Yb(\3P0) atoms,
\begin{eqnarray}
  J &=&
  -\frac{4 \pi \hbar^2 \big(a_S - a_T\big)}{M},
  \label{J-def}
\end{eqnarray}
in which $a_S$ and $a_T$ are the singlet and triplet
scattering lengths.
The exchange interaction is ferromagnetic (antiferromagnetic)
when $a_S - a_T$ is positive (negative). It is crucial to note that
the fact that the Yb(\3P0) is trapped (while the Yb(\1S0) is itinerant 
affects the value of $J$. The
effect of trapping of the Yb(\3P0) impurity on the scattering lengths
is calculated in the Appendix \ref{subsec:CIR}.
It is shown there that  {\it for free Yb atoms}, $a_S > a_T$, and
the exchange interaction is ferromagnetic. On the other hand, 
localization of the Yb(\3P0) atom can modify
the scattering lengths and leads to confinement-induced
resonances (CIR), which change the exchange interaction from
ferromagnetic to antiferromagnetic and vise versa
[see Fig. \ref{Fig-aSaT-eff} for illustration].

\subsection{Kondo Hamiltonian and the Kondo Temperature}
  \label{subsec-HK}

Due to centrifugal barrier, only atoms with $L=0$ interact
with the impurity. Omitting the states  with nonzero $L$,
we write the Hamiltonian of the system as
\begin{equation} \label{HKondo}
  H =
  H_0+H_K,
\end{equation}
where
\begin{eqnarray}
  &&
  H_0 ~=~
  \sum_{n \mu}
  \veps_{n}
  c_{n \mu}^{\dag}
  c_{n \mu},
  \label{H0-def}
  \\
  &&
  H_K ~=~
  J
  \sum_{n n'}
  \Psi_{\g}^{n}(0)
  \Psi_{\g}^{n'}(0)
  \sum_{\mu}
  Z^{\mu\mu}
  c_{n' \mu}^{\dag}
  c_{n \mu}+
  \nonumber \\ && ~~~~~ ~~ +
  J
  \sum_{n n'}
  \Psi_{\g}^{n}(0)
  \Psi_{\g}^{n'}(0)
  \sum_{\mu\neq\mu'}
  X^{\mu\mu'}
  c_{n' \mu'}^{\dag}
  c_{n \mu}.
  \label{HK-def}
\end{eqnarray}
Here $c_{n\mu}$  ($c_{n\mu}^{\dag}$) is the annihilation
(creation) operator for an atom of the Fermi gas, with harmonic
quantum number $n$ and nuclear spin quantum number
$\mu=-\tfrac{5}{2},-\tfrac{3}{2},\ldots,\tfrac{5}{2}$.
$X^{\mu\mu'}=|\mu\rangle\langle\mu'|$ are the Hubbard
operators coupling different degenerate impurity states,
and
\begin{eqnarray*}
  Z^{\mu\mu} &=&
  X^{\mu\mu}-
  \frac{1}{N}
  \sum_{\mu'}
  X^{\mu'\mu'},
  \ \ \ \ \
  N=6.
\end{eqnarray*}

The density of states for the Hamiltonian $H_0$ is,
\begin{eqnarray}
  \rho(\epsilon) &=&
  \sum_{n}
  \Big|
      \Psi_{\g}^{n}(0)
  \Big|^{2}
  \delta(\epsilon - \veps_n)
  \nonumber \\ &\approx&
  \frac{1}{b_{\e}^{3}}~
  \frac{\sqrt{\epsilon}}{\big(2 \hbar \Omega_{\e}\big)^{3/2}}~
  \Theta(\epsilon),
  \label{DOS}
\end{eqnarray}
where $\Theta(\epsilon)$ is the Heaviside theta function.

Within poor man scaling formalism, the dimensionless
coupling $j=J\rho(\epsilon_F)$ satisfies the following
scaling equation \cite{Hewson-book}:
\begin{eqnarray}
  \frac{\partial j}{\partial \ln D}
  &=&
  -N j^2.
  \label{scaling-eq}
\end{eqnarray}
Initially, the bandwidth is $D_0 = T_{\e}$,
$$
  T_{\e} = \frac{\hbar \Omega_{\e}}{k_B},
$$
and the reduced bandwidth $D$ satisfies the inequalities
${D_0}\geq{D}\gg{T}$.
The initial value of $j(D)$, $j(D_0)\equiv{j}_{0}$ is,
\begin{eqnarray}
  j_0 &=&
  \frac{J_0}{b_{\e}^{3}}~
  \frac{\sqrt{\epsilon_F}}{\big(2 \hbar \Omega_{\e}\big)^{2}}
  \nonumber \\ &=&
  -\frac{\pi \big(a_S - a_T\big)}{b_{\e}}~
  \sqrt{\frac{2 \epsilon_F}{\hbar \Omega_{\e}}}.
  \label{j0-dimensionless}
\end{eqnarray}
Eq. (\ref{j0-dimensionless}) shows that the KE exists when
$a_S < a_T$. Scattering lengths for free Yb(\1S0) and Yb(\3P0)
atoms are \cite{Scazza-Yb-3P0-2015}
$a_S = \big(1878 \pm 37\big)~a_B$ and
$a_T = \big(219.7 \pm 2.2\big)~a_B$,
and the exchange interaction is ferromagnetic.
The situation is changed when Yb(\3P0) atom is trapped.
When the ground-state energy of the two atoms is equal to
an energy of the trapped molecule, a confinement-induced
resonance (CIR) occurs \cite{CIR-PhysRevLett-2003, %
CIR-PhysRevLett-2010, CIR-PhysRevLett-2005, %
CIR-PhysRevA-2006}.
In this case, the exchange interaction changes from
ferromagnetic to antiferromagnetic. CIR is considered
in Appendix \ref{subsec:CIR}. It is shown that for
\begin{equation}
  1692 a_B <
  b_{\e} <
  1819 a_B,
  \label{be-res-S-1st}
\end{equation}
$a_S < 0$, and the exchange interaction is antiferromagnetic,
whereas for
\begin{equation}
  212.8 a_B <
  b_{\e} <
  226.7 a_B,
  \label{be-res-T-1st}
\end{equation}
$a_T$ is positive and very large, and the exchange interaction is
antiferromagnetic.

The scaling equation (\ref{scaling-eq}) has the solution,
\begin{eqnarray}
  j(T) &=&
  \frac{1}{N\ln\big(T/T_K\big)},
  \label{scaling-solution}
\end{eqnarray}
where the Kondo temperature (the scaling invariant of the RG
equation) is given by
\begin{eqnarray}
  T_K &=&
  D_0
  \exp
  \bigg(
       -\frac{1}{N j_0}
  \bigg).
  \label{TK}
\end{eqnarray}

\begin{figure}[htb]
\centering
\subfigure[]
{\includegraphics[width=40 mm,angle=0]
  {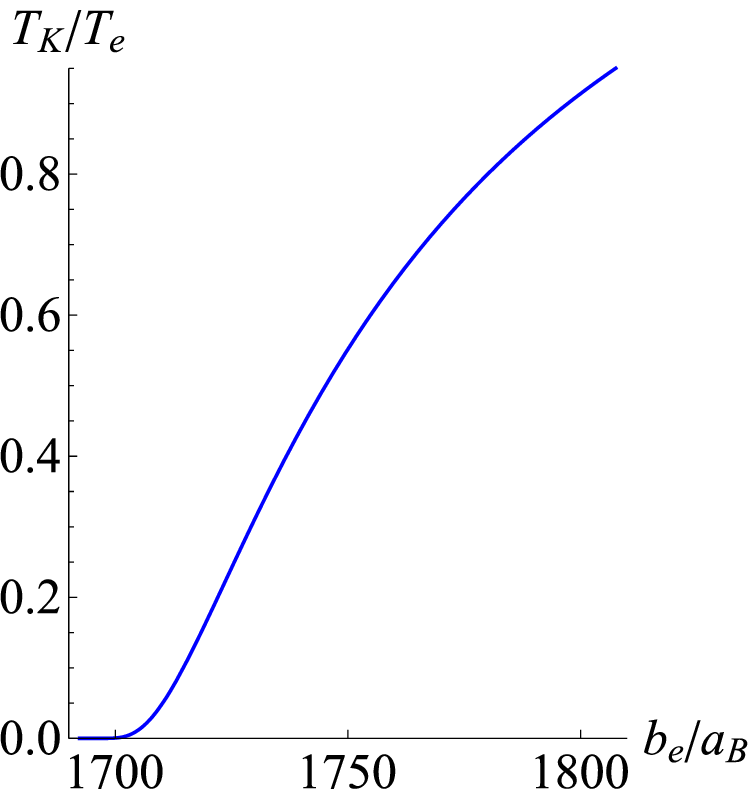}
  \label{Fig-S-1st}}
\subfigure[]
{\includegraphics[width=40 mm,angle=0]
  {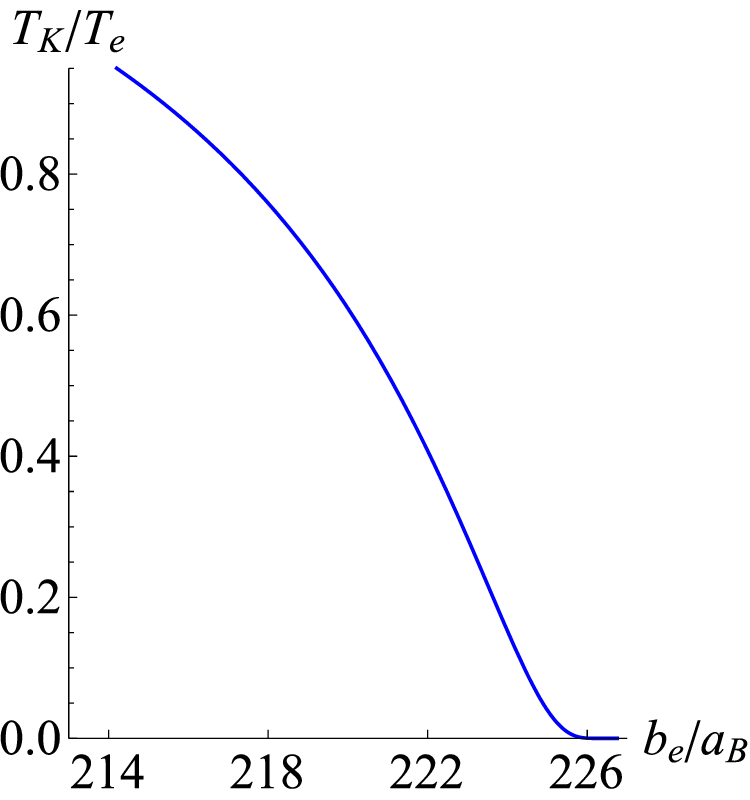}
  \label{Fig-T-1st}}
 \caption{\footnotesize
   Kondo temperature (\ref{TK}) as a function of
   $T_{\e}$ for $D_0 = T_F = T_{\e}$ and
   $b_{\e}$ belonging to
   (a) the interval (\ref{be-res-S-1st}) and
   (b) the interval (\ref{be-res-T-1st}).}
 \label{Fig-TK}
\end{figure}

The Kondo temperature (\ref{TK}) as a function of $T_{\e}$ is shown in
Fig. \ref{Fig-TK} for $D_0 = T_F = T_{\e}$ (purple curves).
It is seen that $T_K$ is large enough and can be measured
in experiment.

\section{Calculation of Observables}
\label{sec-observables}
Having set up the model and the corresponding
Coqblin-Schrieffer Hamiltonian, it is then possible
to predict experimentally measurable observables.
At this stage we are content with presenting a few
thermodynamics quantities appropriate for a system
in thermal equilibrium.  These include the impurity
contributions to the magnetic susceptibility,
specific heat and entropy. Calculations in
the weak coupling regime $T > T_K$ require
different techniques than those in the strong
coupling regime, hence they are presented
separately. Specifically, in the weak coupling
regime one applies the RG formalism, while in
the strong coupling regime the Bethe Ansatz (BA)
analyses is employed.  Both techniques are well
documented and the resulting quantities are
universal functions of $T/T_K$.  Since we have
already  estimated $T_K$, we can use the known
universal expressions for computing and
presenting the pertinent thermodynamic
quantities.

\subsection{Magnetic Susceptibility, specific
            heat and entropy in the Weak Coupling
            Regime}
  \label{sec-suscept-weak}

Since the ytterbium atoms are in a quantum state where
the total electronic angular momentum is zero, the only
contribution to magnetism is due to the nucleus
(the nuclear spin is $5/2$). \\
\underline{Magnetization}:
The impurity contribution to the
magnetization is defined through
the relation \cite{Hewson-book},
\begin{eqnarray}
  {\mathbf{M}}_{\mathrm{imp}} &=&
  g_{\mathrm{Yb}}
  \mu_{\mathrm{n}}
  \Big\{
      \big\langle
          {\mathbf{S}}+
          {\mathbf{s}}
      \big\rangle-
      \big\langle
          {\mathbf{s}}
      \big\rangle_{0}
  \Big\},
\end{eqnarray}
where $\langle\cdots\rangle$ indicates thermal averaging with
respect to the full Hamiltonian $H$, whereas
$\langle\cdots\rangle_{0}$ indicates thermal averaging respect to
$H_0$. $g_{\mathrm{Yb}}=-0.2592$ is the nuclear g-factor of
$^{173}$Yb  \cite{nuclear-moment}, and $\mu_{\mathrm{n}}$ is
the nuclear magneton,
\begin{eqnarray*}
  \mu_{\mathrm{n}} &=&
  \frac{e\hbar}{2m_{\mathrm{p}}c},
\end{eqnarray*}
where $m_{\mathrm{p}}$ is the proton rest mass, and $c$ is
the speed of light. ${\mathbf{S}}$ and ${\mathbf{s}}$ are
the nuclear spin operators for the impurity and the itinerant
atoms, explicitly written as
\begin{eqnarray*}
  &&
  {\mathbf{S}} ~=~
  \sum_{\mu\mu'}
  {\mathbf{t}}_{\mu\mu'}
  X^{\mu\mu'},
  \\
  &&
  {\mathbf{s}} ~=~
  \sum_{{\mathbf{n n}}' \mu\mu'}
  {\mathbf{t}}_{\mu\mu'}
  c_{{\mathbf{n}}\mu}^{\dag}
  c_{{\mathbf{n}}'\mu'},
\end{eqnarray*}
where $\hat{\mathbf{t}}=(\hat{t}^{x},\hat{t}^{y},\hat{t}^{z})$
is a vector of the spin $5/2$ matrices.
In the weak coupling limit, the zero-field magnetic
susceptibility calculated within the poor man's scaling technique
is \cite{Hewson-book},
\begin{eqnarray}
  \chi(T) &=&
  \frac{\chi_0 T_K}{T}~
  \bigg\{
       1-
       \frac{2}{N\ln\big(T/T_K\big)}
  \bigg\},
  \label{suscept-RG}
\end{eqnarray}
where
\begin{eqnarray}
  \chi_0 &=&
  \frac{g_{\mathrm{Yb}}^{2}
        \mu_{\mathrm{n}}^{2}}
       {4 T_K}.
  \label{chi0}
\end{eqnarray}

\begin{figure}[htb]
\centering
   \subfigure[]
   {\includegraphics[width=65 mm,angle=0]
   {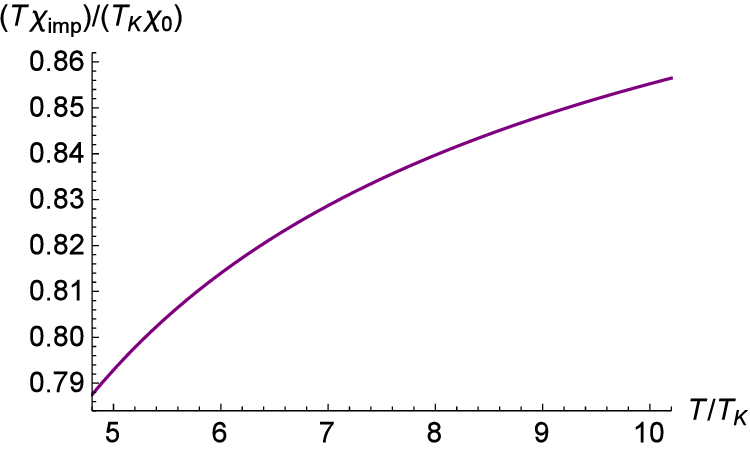}
   \label{Fig-chiT-weak}
   }
   \subfigure[]
   {\includegraphics[width=65 mm,angle=0]
   {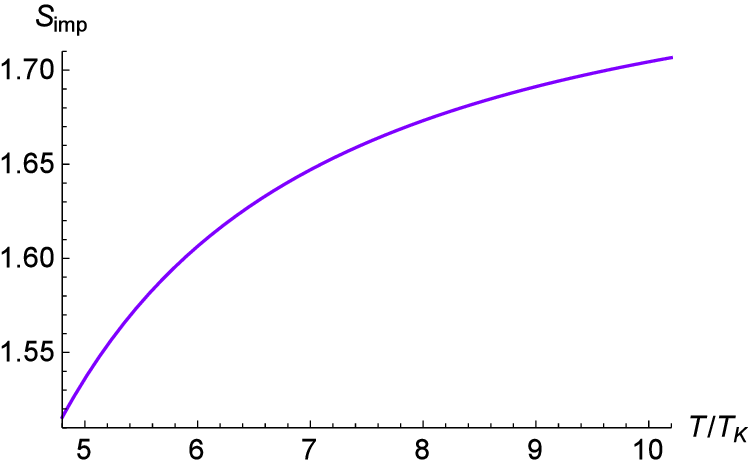}
   \label{Fig-entropy}
   }
   \subfigure[]
   {\includegraphics[width=65 mm,angle=0]
   {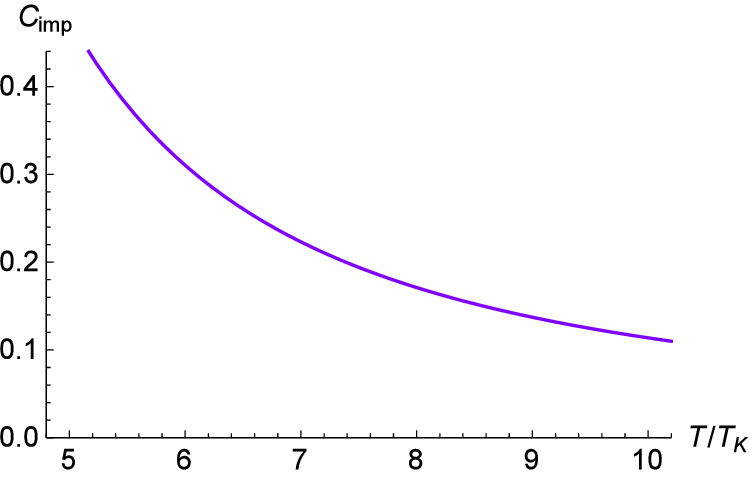}
   \label{Fig-specific-heat}
   }
 \caption{\footnotesize
   ($T\chi(T)$ [panel (a)],
   entropy (\ref{entropy-RG}) [panel (b)] and
   specific heat (\ref{specific-heat-RG}) [panel (c)]
   as functions of temperature $T$ in the weak
   coupling regime ${T}\gg{T_K}$.}
 \label{Fig-chiT-entropy-heat-weak}
\end{figure}

The quantity $T\chi(T)$ are shown in Fig.
\ref{Fig-chiT-weak}. Within the realm of solid
state physics, the mild logarithmic increase of
$\chi(T)$  with decreasing temperature  in the weak
coupling regime of the Coqblin-Schrieffer model,
has been discussed  experimentally and theoretically
a long time ago [see Ref.\cite{Hewson-book}, page 258
Fig.(II)]. It would be of extreme interest to reveal it also
within the realm of cold atom physics. For the isolated
impurity, $T\chi(T)=T_K\chi_0$ is a constant (Curie law).
The fact that $T\chi(T)$ decreases with temperature is
a manifestation of the Kondo interaction of the impurity
with the Fermi gas.
In order to compare the standard SU(2) Kondo model
with the SU($N$) Coqblin-Schrieffer model, we consider
the quantity $X=(T\chi-T_K\chi_0)/(T_K\chi_0)$. For
the SU(2) Kondo effect, $X=1/\ln(T/T_K)$, whereas for
the SU($N$) Coqblin-Schrieffer model
$X=2/(N\ln(T/T_K))$, i.e., the additional factor $2/N$ appears.

\noindent
\underline{Entropy and specific heat:}
Calculations of entropy and specific heat start from the free
energy of the impurity $F_{\mathrm{imp}}=-T\ln(Z/Z_0)$,
where $Z$ is the partition function of the entire system
and $Z_0$ is the partition function of the Fermi gas
without the impurities,
\begin{eqnarray}
  Z =
  {\mathrm{tr}}~
  e^{-\beta H},
  \ \ \ \ \
  Z_0 =
  {\mathrm{tr}}~
  e^{-\beta H_0}
  \label{Z-def}
\end{eqnarray}
The impurity entropy is defined as,
$$
  S_{\mathrm{imp}}=
  -\frac{\partial F_{\mathrm{imp}}}
        {\partial T}.
$$
Poor man's scaling technique which is used in
the weak coupling regime ${T}\gg{T}_{K}$ yield
the following expression for the impurity contribution
to the entropy \cite{Hewson-book}:
\begin{eqnarray}
  S_{\mathrm{imp}} &=&
  \ln\big(N\big)-
  \frac{N^2-1}{N^3}~
  \frac{2\pi^2}
       {3\ln^3(T/T_K)}.
  \label{entropy-RG}
\end{eqnarray}

\noindent
The impurity specific heat
$C_{\mathrm{imp}}=TdS_{\mathrm{imp}}/dT$ is,
\begin{eqnarray}
  C_{\mathrm{imp}} &=&
  \frac{N^2-1}{N^3}~
  \frac{2\pi^2}
       {\ln^4(T/T_K)}.
  \label{specific-heat-RG}
\end{eqnarray}

The entropy (\ref{entropy-RG}) and
the specific heat (\ref{specific-heat-RG}) as
functions of temperature are displayed in Figs.
\ref{Fig-entropy} and \ref{Fig-specific-heat}.
Kondo effect results in reducing of the entropy
with decreasing temperature, whereas
$C_{\mathrm{imp}}$ increases when temperature
decreases. This is the manifestation of the Kondo
effect. Note that for the standard SU(2) KE (which
is the case $N=2$), both $S_{\mathrm{imp}}-\ln(2)$
and $C_{\mathrm{imp}}$ are proportional to the factor
$3/8$, whereas for the SU($N$) Coqblin-Schrieffer
model, the factor $(N^2-1)/N^3$ appears.

\subsection{Magnetization, specific heat and
           entropy in the strong coupling regime}
  \label{sec-strong}

For $T<T_K$, a non-perturbative method should be
employed for calculating observables. This is worked
out in Ref. \cite{Rajan-prl-83}, where the BA
was applied for studying the Coqblin-Schrieffer model
at low temperature. Here we apply the formalism derived
therein for calculating the pertinent observables in thermal
equilibrium. The general structure and behaviour of
these quantities is expressed as universal functions of 
$T/T_K$. In particular, the magnetic susceptibility
$\chi_{\mathrm{imp}}$, the ratio $S_{\mathrm{imp}}/T$
between the entropy and temperature and
the ratio $C_{\mathrm{imp}}/T$ between
the specific heat and the temperature are characterized
by a {\textit{finite temperature peak}} that becomes more
dominant at larger $N$. This is the main difference
between the standard SU(2) Kondo model and
SU($N$) Coqblin-Schrieffer model: For the SU(2) KE, each
one of these three quantities displays a {\textit{zero temperature
peak}} \cite{Rajan-prl-83,Hewson-book}.

The contribution of the  impurity to the free energy at
a given magnetic field $B$ (and for $T<T_K$) reads
\cite{Rajan-prl-83},
\begin{eqnarray}
  F_{\mathrm{imp}} &=&
  -T
  \sum_{\mu}
  \int{d\epsilon}~
  \rho_{\mathrm{sc}}\big(\epsilon-\mu \Delta_B\big)
  \ln
  \Big(
      1+e^{-\epsilon/T}
  \Big)+
  \nonumber \\ && +
  T
  \int{d\epsilon}~
  \rho_{\mathrm{sc}}\big(\epsilon\big)
  \ln
  \Big(
      1+e^{-\epsilon/T}
  \Big),
  \label{free-energy-BA}
\end{eqnarray}
where
$$
  \Delta_B ~=~
  g_{\mathrm{Yb}} \mu_n B,
$$
the energy $\epsilon$ is measured with respect to
the Fermi energy.
Here $\rho_{\mathrm{sc}}(\epsilon)$ is the density of state
(DOS) of fermions calculated in the strong coupling limit.
At zero temperature, there is a peak in the DOS of width
of order $T_K$ near the Fermi energy \cite{Hewson-book}.
This peak is calculated in the framework of slave boson
mean field theory \cite{Hewson-book,KA-prb14},
\begin{eqnarray}
  \rho_{\mathrm{sc}}(\epsilon) &=&
  \frac{1}{2}
  \sum_{\nu=\pm1}
  \frac{\frac{\pi T_K}{N}}
       {\big(\epsilon-\nu\epsilon_N\big)^2+
        \left(\frac{\pi T_K}{N}\right)^2},
  \label{DOS-MFSBA}
\end{eqnarray}
where $\epsilon_N=T_K\cos(\pi/N)$. Here we take into
account that the electron excitations and hole excitations
contribute equally to the free energy \cite{Rajan-prl-83},
and  take $g(\epsilon)=g(-\epsilon)$.

\noindent
\underline{Magnetic susceptibility:} The zero field impurity
magnetic susceptibility $\chi_{\mathrm{imp}}$, defined as
$$
  \chi_{\mathrm{imp}} ~=~
  -\bigg(
       \frac{\partial^2 F_{\mathrm{imp}}}
            {\partial B^2}
  \bigg)_{B\to0},
$$
is given by \cite{Rajan-prl-83},
\begin{eqnarray}
  \chi_{\mathrm{imp}} &=&
  \frac{\chi_0 T_K}
       {12T}~
  N\big(N^2-1\big)
  \int
  \frac{\rho_{\mathrm{sc}}(\epsilon)d\epsilon}
       {\cosh^2\left(\frac{\epsilon}{2T}\right)},
  \label{suscept-SBMFA}
\end{eqnarray}
where $\chi_0$ is given by Eq. (\ref{chi0}).
\begin{figure}[htb]
\centering
  \subfigure[]
  {\includegraphics[width=65 mm,angle=0]
   {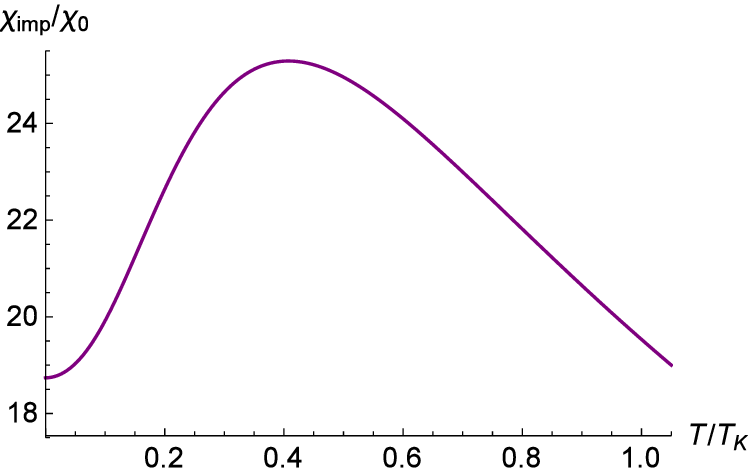}
   \label{Fig-chi-BA}
   }
   \subfigure[]
   {\includegraphics[width=65 mm,angle=0]
   {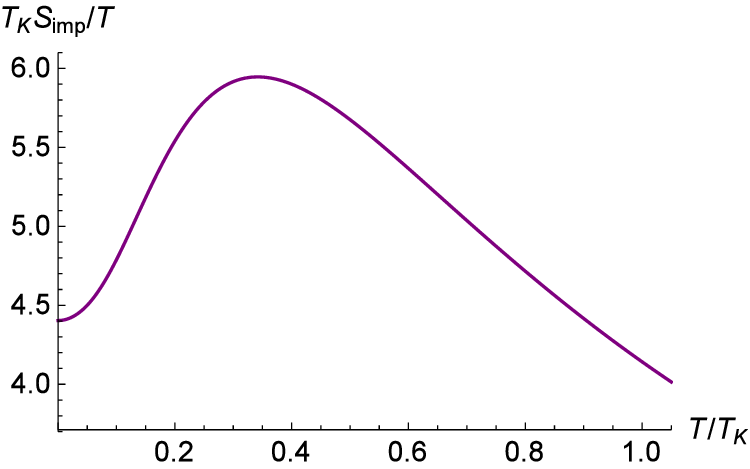}
   \label{Fig-entropy-BA}
   }
   \subfigure[]
   {\includegraphics[width=65 mm,angle=0]
   {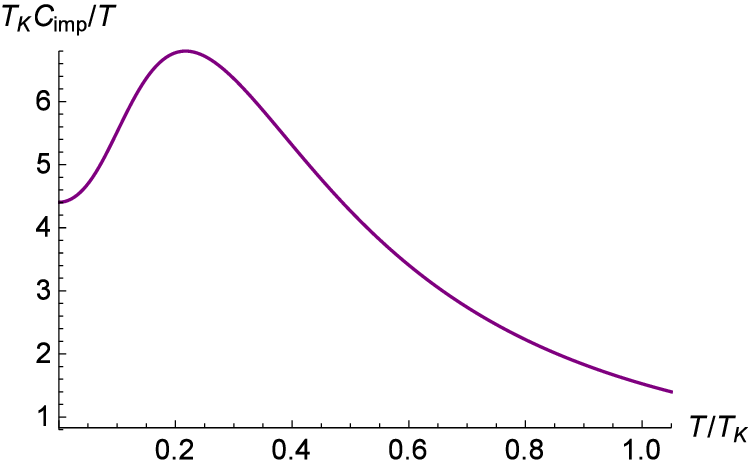}
   \label{Fig-specific-heat-BA}
   }
 \caption{\footnotesize
    (Magnetic susceptibility (\ref{suscept-SBMFA}) [panel (a)],
   $S_{\mathrm{imp}}/T$ [panel (b)] and
   $TC_{\mathrm{imp}}(T)$ [panel (c)]
   as functions of temperature in the strong coupling
   regime $T<T_K$}.
 \label{Fig-chi-entropy-heat-BA}
\end{figure}
The magnetic susceptibility (\ref{suscept-SBMFA})
is shown in Fig. \ref{Fig-chi-BA}.

\noindent
\underline{Entropy:} Differentiating the free energy
(\ref{free-energy-BA}) and letting the magnetic field
$B \to 0$, we obtain an expression for the impurity
entropy,
\begin{eqnarray}
  S_{\mathrm{imp}} &=&
  (N-1)
  \int{d\epsilon}~
  \rho_{\mathrm{sc}}\big(\epsilon\big)
  \ln
  \Big(
      1+e^{-\epsilon/T}
  \Big)+
  \nonumber \\ && +
  \frac{N-1}{T}
  \int{d\epsilon}~
  \epsilon
  \rho_{\mathrm{sc}}\big(\epsilon\big)
  f(\epsilon),
  \label{entropy-BA}
\end{eqnarray}
where $f(\epsilon)$ is the Fermi-Dirac distribution,
$$
  f(\epsilon) ~=~
  \frac{1}{1+e^{\epsilon/T}}.
$$

\noindent
\underline{Specific heat:} Differentiating the entropy,
we get specific heat of the impurity \cite{Rajan-prl-83},
\begin{eqnarray}
  C_{\mathrm{imp}} &=&
  \big(N-1\big)
  \int
  \Big(
       \frac{\epsilon}{2T}
  \Big)^{2}~
  \frac{\rho_{\mathrm{sc}}(\epsilon)d\epsilon}
       {\cosh^2\left(\frac{\epsilon}{2T}\right)}.
  \label{specific-heat-BA}
\end{eqnarray}

The functions $S_{\mathrm{imp}}/T$ and $C_{\mathrm{imp}}/T$ are
shown in Figs. \ref{Fig-entropy-BA} and \ref{Fig-specific-heat-BA}.

\section{Conclusions}
  \label{sec-conclusions}
We have studied the feasibility of realizing the Coqblin-Schrieffer
model in cold $^{173}$Yb atoms. The peculiarities of this framework
are as follows:
1) The same atoms are used as itinerant fermions and impurity
atom, the only difference is that the former is in an excited state
$^3$P$_0$ and the latter is in the atomic ground-state
$^1$S$_0$.
2) For both ground and excited states, the electronic total
angular momentum is zero.
3) Therefore the ensuing exchange interaction is indirect
and proceeds through virtual ionic states
$$
  [(6s^2),(6s6p)] \to
  [(6s^26p)^-,(6s)^+] \to
  [(6s6p),(6s^2)].
$$
The corresponding (positive) exchange energy
between the localized and itinerant ytterbium
atoms  is calculated using reasonable models
of atomic wave functions and experimental data
for scattering lengths obtained in Ref.~\cite{6}.
It  is then incorporated within a Coqblin-Schrieffer
Hamiltonian, and the Kondo temperature is
estimated to be $T_K=~0.16\sim0.31~T_F$ for
$T_F=~50\sim300$~nK. These conditions are
favourable for the Kondo effect to be observed
in cold fermionic ytterbium laboratories. Using
renormalization group analysis, we calculated
the magnetic susceptibility, entropy and specific
heat of the impurity  in the weak coupling regime,
$T\gg{T}_{K}$. The temperature behaviour of
these two quantities is in (qualitative) agreement
with calculations carried out in heavy fermion
systems, specifically for the $\Gamma_8$
quartet $S=\frac{3}{2}$ in a system of Ce impurity
immersed in a LaB metal under cubic crystal
field \cite{Schlottmann}.

In the second step, we used the machinery of
the Bethe Ansatz formalism\cite{BA} for
the calculation of the impurity  contribution
to the magnetic susceptibility,  entropy and
specific heat with the specific parameters
pertaining to our Yb system (such as $J$,
$T_K$, $N$, $g_{\mathrm{Yb}}$ and
$\mu_n$). These results should consist of
a reference starting point for relevant experiments.

\noindent
{\bf Acknowledgements}\\ 
Invaluable correspondence and discussion
with Leonid Isaev are highly appreciated.
The research of IK, TK and YA is partially
supported by the Israel Science Foundation
(ISF) under Grant 400/2012. G.B.J acknowledges
financial support from the Hong Kong Research
Grants Council (Project No.ECS26300014).

\appendix

\section{Derivation of the Exchange Interaction}
 \label{append-exchange}

In this appendix we derive an expression
for the exchange interaction between two
atoms of $^{173}$Yb. One of them
(numbered 1) is the long-lived $^{3}$P$_{0}$ excited 
 state with nuclear spin $\mu$,
and the other one (numbered 2) is the $^{1}$S$_{0}$ ground
 state with nuclear spin $\mu'$.
Each atom is considered as composed of
an inert core (charge (+2) closed shell rigid ion)
and two valence electrons, as illustrated in
Fig. \ref{Fig-atoms}. The electron configuration
of the excited state $^{3}$P$_{0}$  is $6s6p$,
while that of the ground state $^{1}$S$_{0}$ is $6s^2$.
The positions of the ions are denoted as
$\mbfR_{\e}$ and $\mbfR_{\g}$.

\subsection{Indirect Exchange Interaction}
  \label{append-hybridization}

\begin{figure}[htb]
\centering
\subfigure[]
  {\includegraphics[width=55 mm,angle=0]
   {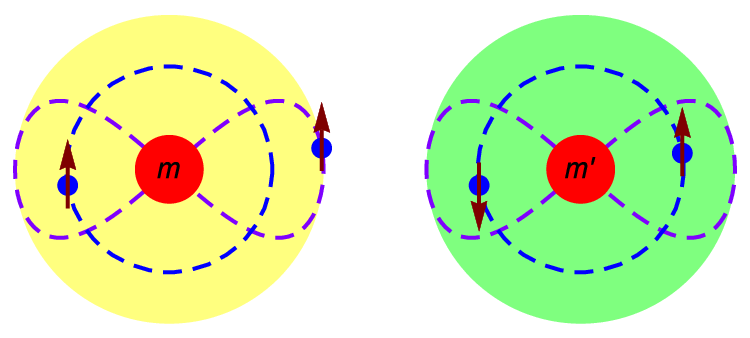}}
\subfigure[]
  {\includegraphics[width=55 mm,angle=0]
   {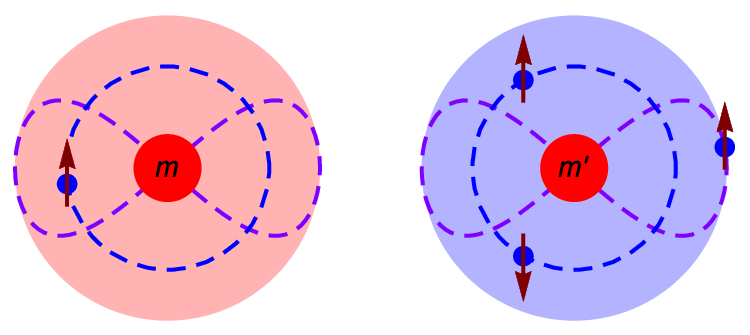}}
\subfigure[]
  {\includegraphics[width=55 mm,angle=0]
   {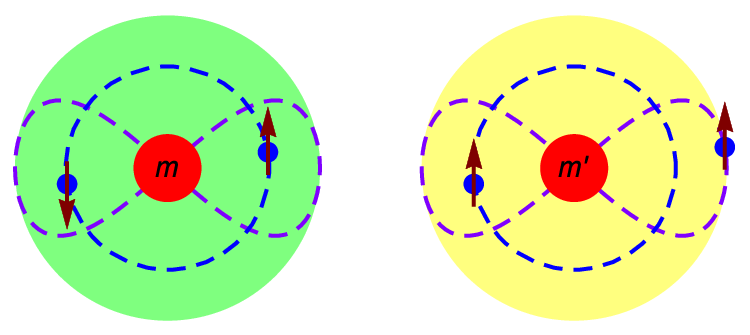}}
 \caption{Illustration of exchange
   interaction between ytterbium atoms.
   Panel (a): Initial quantum state - the first atom
   (numbered 1) is in the meta-stable state (light yellow disk)
   and the second one (numbered 2) is in the ground state
   (light green disk);
  panel (b): virtual state - the first atom is positively ionized
  (light red disk), and the second one is negatively charged
  (light blue disk);
  panel (c): final state - the first atom is in the ground
  state and the other one is in the meta-stable state.
  For all the panels, arrows denote the electronic spin,
  $m$ or $m'$ is nuclear spin of the first or second
  atom.}
 \label{Fig-GX-II-XG}
\end{figure}

In the present subsection we consider tunnelling of an electron from one atom
to another which turns two neutral atoms into two ions or
{\it vice-versa}, two ions into neutral atoms, as illustrated in Fig. \ref{Fig-GX-II-XG}.
Explicitly, a 6$p$ electron tunnels from the atom
in the excited state to the atom in the ground state.
As a result, we have two ions with parallel electronic
orbital moments [Fig. \ref{Fig-GX-II-XG}(b)]. Then
one electron from the 6$s$ orbital tunnels from
the negatively charged ion to the 6$s$ orbital of
the positively charged ion. The net outcome is that
the atoms ``exchange their identity'' specified by
their electronic quantum states: one atom transforms
from the ground state to the excited state, whereas the other
atom transforms from the excited state to the ground state.

The Kohn-Sham scheme of density functional theory
\cite{e-density-1996, e-density-ChemPhysLett-1997, e-density-JChemPhys-2002,%
e-density-Pollet-IntJQuantumChem-2003, e-density-Savin-IntJQuantumChem-2003,
e-density-PhysRevA-2004, e-density-JChemPhys-2005}
enables one to describe 
the long-range part of the potential created by
an electron cloud as,
\begin{eqnarray}
  \Phi(R) &=&
  -\frac{2e}{R}~
  {\mathrm{erf}}\big(\mu R\big),
  \label{Phi-long-range}
\end{eqnarray}
where $R$ is the distance from the nucleus and $1/\mu$ represents
the distance beyond which the interaction reduces to the usual
Coulomb long-range tail.
The coefficient 2 in the numerator indicates 2 outer electrons
of the Yb atom [namely, $6s^2$ or $6s6p$ electrons of the Yb(\1S0)
or Yb(\3P0) atom]. It is assumed that $\mu$ is the same 
for the $6s$ and $6p$ electrons.  It is used below as a parameter
for fitting the scattering lengths.

The Coulomb interaction between an electron and Yb(\1S0) and Yb(\3P0) atoms
separated by distance $R$ can be written as
\begin{eqnarray}
  &&
  V(\mbfr,\mbfR) =
  -\frac{2 e^2}{\big|\mbfr - \mbfR/2\big|}~
  \Big[
      1 - {\mathrm{erf}}\big(\mu |\mbfr - \mbfR/2|\big)
  \Big]
  \nonumber \\ && ~~~~~ ~~~~~ -
  \frac{2 e^2}{\big|\mbfr + \mbfR/2\big|}~
  \Big[
      1 - {\mathrm{erf}}\big(\mu |\mbfr + \mbfR/2|\big)
  \Big],
  \label{V-erf}
\end{eqnarray}
where $\mbfr$ is the position of the electron, and $\pm \mbfR/2$
are the positions of the atoms.

The tunnelling rate for an electron between the atoms can be calculated using the WKB approximation,
\begin{eqnarray}
  t_{\nu}(R) &=&
  \epsilon_{\nu}~
  e^{-2 \gamma_{\nu}(R)},
  \nonumber
  \\
  \gamma_{\nu}(R) &=&
  \int\limits_{-z_0(\epsilon_{\nu})}^{z_0(\epsilon_{\nu})}
  \frac{\sqrt{2 m_e}~dz}{\sqrt{\hbar^2 \big(V(z,R) - \epsilon_{\nu}\big)}},
  \label{tunneling-rate-WKB}
\end{eqnarray}
where $\nu = s,p$ for the $6s$ and $6p$ electrons,
$m_e$ is the mass of electron, $V(z,R)$ is
the potential energy (\ref{V-erf}) for $\mbfr = (0,0,z)$ and
$\mbfR = (0,0,R)$, and $\pm z_0(\epsilon_{\nu})$ are classical
turning points. The single-electron energies are
$\epsilon_s = 6.2542$~eV and $\epsilon_p = 4.1107$~eV.

We calculate $t_{\nu}(R)$ numerically for different $R$ and
found that it can be approximated by the simple function,
\begin{eqnarray}
  t_{\nu}(R) &=&
  t_{\nu}^{(0)}~
  e^{- 2 A_{\nu}^{(1)} (R - R_0)},
  \label{tunnel-fit}
\end{eqnarray}
where $R_0 = 4$~{\AA}, $t_{\nu}^{(0)} = \epsilon_{\nu} e^{-2 A_{\nu}^{(0)}}$,
and $A_{\nu}^{(0)}$ and $A_{\nu}^{(1)}$ are given by
\begin{eqnarray*}
  &&
  A_{s}^{(0)} = 1.579,
  \ \ \ \ \
  A_{s}^{(1)} = 1.284~{\text{\AA}}^{-1},
  \nonumber
  \\
  &&
  A_{p}^{(0)} = 0.943,
  \ \ \ \ \
  A_{p}^{(1)} = 1.044~{\text{\AA}}^{-1}.
\end{eqnarray*}
Explicitly, $t_{s}^{(0)} = 0.266$~eV and
$t_{p}^{(0)} = 0.248$~eV.
Accordingly, the exchange interaction due to tunnelling of the $6s$ and $6p$ electrons is
\begin{eqnarray}
  g(R) &=&
  \frac{2 t_s(R) t_p(R)}{\Delta \veps},
  \label{g-def}
\end{eqnarray}
where
\begin{eqnarray*}
  \Delta\veps &=&
  \veps_{\mathrm{ion}}-
  \veps_{\mathrm{ea}}+
  \epsilon_1-
  \epsilon_2
  ~=~
  4.4107~{\mathrm{eV}},
\end{eqnarray*}
where $\veps_{\mathrm{ion}}=6.2542$~eV is the ionization
energy \cite{Yb-spectrum-78} and
$\veps_{\mathrm{ea}}=-0.3$~{eV} is the electron affinity
\cite{Electron-afinity-PhRep04} of ytterbium.

\subsection{Semiclassical Wave Function in Atomic Scattering}
  \label{subsection-WKB}

Here we solve Schr\"odinger equation
(\ref{eq-Schrodinger-2atoms-small-R})
in the framework of the semiclassical approximation
\cite{Scatt-Length-WKB-PRA93,vdW-Yb-PRA-08}.
The potential $W(R)$, Eq. (\ref{vdW-def}),
depends just on the distance $R$
from the impurity. Therefore, the orbital momentum $L$
and its projection $m$ on the axis $z$ are good quantum
numbers. Because of the centrifugal barrier, atoms with
nonzero $L$ cannot approach close one to another.
Therefore we restrict ourselves by considering just
the s-wave (i.e., the wave with $L=0$).
Represent the wave function
$\Psi_{r}(\mbfR)\equiv\Psi_{n}(\mbfR)$ as
\begin{eqnarray}
  \Psi_{n}(\mbfR) &=&
  \frac{\psi_{n}(R)}{\sqrt{4\pi}~R}.
  \label{WF-partial-append}
\end{eqnarray}
The radial wave function $\psi_{n}(R)$ satisfies the 1D
Schr\"odinger equation,
\begin{eqnarray}
  \psi''_{n}(R)+
  \frac{M}{\hbar^2}~
  W(R)
  \psi_{n}(R)
  = 0.
  \label{eq-Schrodinger-1D-append}
\end{eqnarray}
To solve this equation it is useful to employ different approximations 
in several corresponding intervals as defined below. 
To this end, we underline the following
constraints on the parameters $R$: $r_0$, $b_0$ and $\lambda$ as follows: 
\begin{itemize}
\item $r_0$ is determined from the equation $W(r_0)=0$.
      The classical mechanics allows motion of the zero-energy
      particle in the interval $R>r_0$.

\item $b_0$ is constrained by the inequality,
      $$
        \bigg|
             \frac{\sigma^6}{b_0^6}-
             \frac{C_8}{C_6 b_0^2}
        \bigg|
        ~\ll~ 1.
      $$
      For ${R}\geq{b}_{0}$, we can approximate
      $W(R)\approx-C_6/R^6$. Practically, we take $b_0\approx10$~{\AA}
      [see Fig. \ref{Fig-vdW}].

\item $\lambda=(MC_6/\hbar^2)^{1/4}=89.97$~{\AA}. In principle,
      the WKB approximation
      can be used for for $R\ll\lambda$. 
\end{itemize}
A brief list of approximations per intervals 
is as follows (see details below): 
For the interval $r_0<R\ll\lambda$, we can apply the WKB
approximation to solve the Schr\"odinger equation
(\ref{Schrodinger-eq-relative}). For the interval
$b_0<R<\lambda$, we can approximate $W(R)$ by
$-C_6/R^6$ and solve eq. (\ref{Schrodinger-eq-relative}).
The interval $R<r_0$ corresponds to classically
forbidden region where the wave function decays exponentially. 
In the following discussions, we find the wave function
within each interval. The intervals $r_0<R\ll\lambda$ and
$b_0<R<\lambda$ overlap one with another since
there is a wide interval $b_0<R\ll\lambda$ where
both the WKB approximation and the approximation
$W(R)\approx-C_6/R^6$ are valid. Therefore, within
this interval both the approaches should give
the same solution. We use this condition as as
a connection condition for the solutions within
two overlapping intervals [see eqs.
(\ref{psi-b0<R<lambda-append}) and
(\ref{subeqs-psi-AB-R-small-append}) below].

\noindent
{\textbf{1}}. {\underline{Interval $r_0<R\ll\lambda$}}:
In order to solve equation (\ref{Schrodinger-eq-relative}),
we apply the WKB approximation with quantum corrections
\cite{vdW-Yb-PRA-08}. The wave function within
this approximation is,
\begin{eqnarray}
  \psi_{n}^{(1)}(R) &=&
  \frac{A_{1n}}{\sqrt{K(R)}}~
  \sin
  \Big(
      \Phi_{r}(R)+
      \frac{\pi}{4}
  \Big),
  \label{psi-R0<R<b0-append}
\end{eqnarray}
where $A_{1n}$ is unknown constant,
\begin{eqnarray}
  \Phi_{r}(R) &=&
  \int\limits_{r_0}^{R}
  K(R')dR',
  \label{Phi-def-append}
  \\
  K(R) &=&
  \frac{1}{\hbar}~
  \sqrt{-M W(R)}.
  \label{K-def-append}
\end{eqnarray}
Here the phase $\pi/4$ takes into account connection
of $\psi_{n}^{(1)}(R)$ with exponentially decaying
solution in the classically forbidden interval $R<r_0$
[see Ref. \cite{Scatt-Length-WKB-PRA93,vdW-Yb-PRA-08}].

When $R>b_0$, we can write eq. (\ref{Phi-def-append}) as,
\begin{eqnarray}
  \Phi_{r}(R) =
  \int\limits_{r_0}^{\infty}
  K(R')~dR'-
  \int\limits_{R}^{\infty}
  K(R')~
  dR'.
  \label{Phi(R)-for-long-R-append}
\end{eqnarray}
For any $R\geq{b}_{0}$, $K(R)$ can be approximated by
$K_{0}(R)$ given by the equation,
$$
  K_{0}(R) ~=~
  \frac{\sqrt{M C_6}}{\hbar R^3} ~=~
  \frac{\lambda^2}{R^3}.
$$
Then the second integral on the right hand side of
eq. (\ref{Phi(R)-for-long-R-append}) can be performed analytically
and gives,
$$
  \int\limits_{R}^{\infty}
  \frac{\lambda^2~dR'}{\big(R'\big)^3}
  ~=~
  \frac{\lambda^2}{2R^2}.
$$
Taking into account that the first term on the right hand
side of eq. (\ref{Phi(R)-for-long-R-append}) is
$\Phi_w$, eq. (\ref{Phi-w}), we can write
\begin{eqnarray}
  \Phi_{r}(R) =
  \Phi_w-
  \frac{\lambda^2}{2R^2}.
  \label{Phi(b0)-append}
\end{eqnarray}

Then $\psi_{n}^{(1)}(R)$ for $R>b_0$ takes the form,
\begin{eqnarray}
  \psi_{n}^{(1)}(R) =
  \frac{A_{1n} R^{3/2}}{\lambda}~
  \sin
  \Big(
      \Phi_{w}-
      \frac{\lambda^2}{2R^2}+
      \frac{\pi}{4}
  \Big).
  \label{psi-b0<R<lambda-append}
\end{eqnarray}

\noindent
{\textbf{2}}. {\underline{Interval $R>b_0$}}:
Within this interval, we can approximate the potential energy
by $W(R)\approx-C_6/R^6$ and write the Schr\"odinger
equation (\ref{eq-Schrodinger-1D-append}) in the form,
\begin{eqnarray}
  \lambda^2~
  \frac{d^2\psi_{n}^{(2)}(R)}{d R^2}+
  \frac{\lambda^6}{R^6}~
  \psi_{n}^{(2)}(R)
  &=& 0,
  \label{eq-Schrodinger-1D-2nd-interval-append}
\end{eqnarray}
where $\lambda$ is given by eq. (\ref{lambda-def}).

General solution of eq. (\ref{eq-Schrodinger-1D-2nd-interval-append})
is,
\begin{eqnarray}
  \psi_{n}^{(2)}(R) &=&
  A_{2n}~
  \tilde\psi_{2A}(R)+
  B_{2n}~
  \tilde\psi_{2B}(R),
  \label{psi-b0<R<rk-append}
\end{eqnarray}
where $\tilde\psi_{2A}(R)$ and $\tilde\psi_{2B}(R)$ are two
linearly independent solutions of eq.
(\ref{eq-Schrodinger-1D-2nd-interval-append}),
\begin{subequations}
\begin{eqnarray}
  \tilde\psi_{2A}(R) &=&
  \sqrt{\frac{2R}{\lambda}}~
  J_{1/4}
  \bigg(
       \frac{\lambda^2}{2R^2}
  \bigg),
  \label{psi-A-append}
  \\
  \tilde\psi_{2B}(R) &=&
  \sqrt{\frac{2R}{\lambda}}~
  J_{-1/4}
  \bigg(
       \frac{\lambda^2}{2R^2}
  \bigg),
  \label{psi-B-append}
\end{eqnarray}
  \label{subeqs-psi-AB-append}
\end{subequations}
$A_{2n}$ and $B_{2n}$ are unknown constants.

When $R\gtrsim\lambda$, the asymptotic expressions for
$\tilde\psi_{2A}(R)$ and $\tilde\psi_{2B}(R)$ are,
\begin{subequations}
\begin{eqnarray}
  \tilde\psi_{2A}(R) =
  \frac{1}{\Gamma(5/4)}~
  \Bigg\{
       1-
       \frac{\lambda^4}{20R^4}+
       O\bigg(\frac{\lambda^8}{R^8}\bigg)
  \Bigg\},
  \label{psi-A-R-large-append}
  \\
  \tilde\psi_{2B}(R) =
  \frac{2R/\lambda}{\Gamma(3/4)}~
  \Bigg\{
       1-
       \frac{\lambda^4}{12R^4}+
       O\bigg(\frac{\lambda^8}{R^8}\bigg)
  \Bigg\}.
  \label{psi-B-R-large-append}
\end{eqnarray}
  \label{subeqs-psi-AB-R-large-append}
\end{subequations}
For $R\ll\lambda$, the asymptotic expressions are,
\begin{subequations}
\begin{eqnarray}
  \tilde\psi_{2A}(R) =
  \pi~
  \bigg(
       \frac{2R}{\pi\lambda}
  \bigg)^{3/2}
  \sin
  \bigg(
       \frac{\lambda^2}{2R^2}+
       \frac{\pi}{8}
  \bigg),
  \label{psi-A-R-small-append}
  \\
  \tilde\psi_{2B}(R) =
  \pi~
  \bigg(
       \frac{2R}{\pi\lambda}
  \bigg)^{3/2}
  \cos
  \bigg(
       \frac{\lambda^2}{2R^2}-
       \frac{\pi}{8}
  \bigg).
  \label{psi-B-R-small-append}
\end{eqnarray}
  \label{subeqs-psi-AB-R-small-append}
\end{subequations}

The functions $\tilde\psi_{2A}(R)$ and $\tilde\psi_{2B}(R)$,
eq. (\ref{subeqs-psi-AB-append}), and their asymptotes
(\ref{subeqs-psi-AB-R-large-append}) are shown in Fig.
\ref{Fig-WF-Bessel}, solid and dashed lines. It is seen
that for $R>\lambda$, the functions $\tilde\psi_{2A}(R)$
and $\tilde\psi_{2B}(R)$ are well approximated by their
asymptotic expressions \cite{Scatt-Length-WKB-PRA93}.

There is a large interval $b_0<R\ll\lambda$, where we can
approximate $W(R)$ by $-C_6/R^6$ and apply the WKB
approximation. Therefore, we can apply the following
connection conditions: For any $R$ within the interval
$b_0<R\ll\lambda$, the equality
$\psi_{n}^{(1)}(R)=\psi_{n}^{(2)}(R)$ is valid.
This conditions gives,
\begin{subequations}
\begin{eqnarray}
  A_{2n}
  &=&
  -A_{1n}~
  \frac{\sqrt{\pi \lambda}}{2}~
  \cos
  \Big(
      \Phi_w+
      \frac{\pi}{8}
  \Big),
  \label{A2-vs-A1-append}
  \\
  B_{2n}
  &=&
  A_{1n}~
  \frac{\sqrt{\pi \lambda}}{2}~
  \sin
  \Big(
      \Phi_w+
      \frac{3\pi}{8}
  \Big).
  \label{B2-vs-A1-append}
\end{eqnarray}
  \label{subeqs-A2-B2-vs-A1-append}
\end{subequations}

Taking into account eqs. (\ref{subeqs-A2-B2-vs-A1-append}) and
(\ref{subeqs-psi-AB-R-large-append}), we can write the asymptote
of the wave function $\psi_{n}^{(2)}(R)$, eq. (\ref{psi-b0<R<rk-append}),
as $\psi_{n}^{(2)}(R)\propto{R-a_w}$, with the scattering
length $a_w$ given by eq. (\ref{scattering-length}) [see
Refs. \cite{Scatt-Length-WKB-PRA93,vdW-Yb-PRA-08}].

\noindent
{\textbf{3}}. {\underline{Interval $R\gtrsim\lambda$}}:
Within this interval, the wave function is given by
eq. (\ref{WF-g-3D}). However, it is convenient to
introduce a radial wave function $\psi_{3n}(R)$
similar to eq. (\ref{WF-partial-append}),
\begin{eqnarray}
  &&
  \Psi_{\g}^{n00}(\mbfR) ~=~
  \frac{\psi_{n}^{(3)}(R)}
       {\sqrt{4 \pi}~R},
  \nonumber
  \\
  &&
  \psi_{n}^{(3)}(R) ~=~
  \frac{2^{3/4} \sqrt{k_n}}{\sqrt{\pi}~b_{\g}}~
  \big(
      R-a_w
  \big),
  \label{psi3-def-append}
\end{eqnarray}
where
\begin{eqnarray}
  k_n ~=~
  \frac{2\sqrt{n}}{b_{\g}},
  \label{kn-def-append}
\end{eqnarray}
and $b_{\g}$ is the harmonic length (\ref{a-g-omega-g-def}).
Here we take into account that $k_n\lambda\ll1$ and
$k_n a_w \ll 1$, and approximate $\psi_{3}(R)$ for
$R\gtrsim\lambda$ by a linear function.

The wave function $\psi_n(R)$ and its
derivative $\psi'_{n}(R)$ are continues at $R=\lambda$.
These conditions give
\begin{eqnarray}
  A_{1n} &=&
  \frac{2~
        \sqrt{k_n \lambda}}
       {\pi  b_{\g}}~
  \Gamma\bigg(\frac{3}{4}\bigg)~
  \sqrt{1+\Big(\frac{a_w-\bar{a}}{\bar{a}}\Big)^{2}}.
  \label{A1-res-append}
\end{eqnarray}

\subsection{Comparison of Our Calculations with the Results
          of Ref. \cite{Scazza-Yb-3P0-2015}}
  \label{subsec-comparison}

In Ref. \cite{6}, the authors report measurement 
of scattering lengths for
two ytterbium atoms in the ``singlet'' and ``triplet'' two particle
states [i.e., two particle states with symmetric and antisymmetric
spatial wave function]. Explicitly, they are,
\begin{eqnarray}
  a_S &=&
  \big(1878 \pm 37\big)~a_B,
  \label{aS-experiment}
  \\
  a_T &=&
  \big(219.7 \pm 2.2\big)~a_B.
  \label{aT-experiment}
\end{eqnarray}
In order to compare our results with the measurements of Ref. \cite{6},
we consider scattering of ytterbium atoms in the ground
state with a localized impurity with taking into account
van der Waals and exchange interaction. The van der Waals
interaction between the ytterbium atoms is given by
eq. (\ref{vdW-def}) [see also Ref. \cite{vdW-Yb-PRA-08}].
The exchange interaction between two atoms separated by
distance $R$ is given by Eq. (\ref{g-def}).
Recall that positive $g(R)$ means that the corresponding exchange
interaction  is anti-ferromagnetic.

\begin{figure}[htb]
\centering
\includegraphics[width=65 mm,angle=0]
   {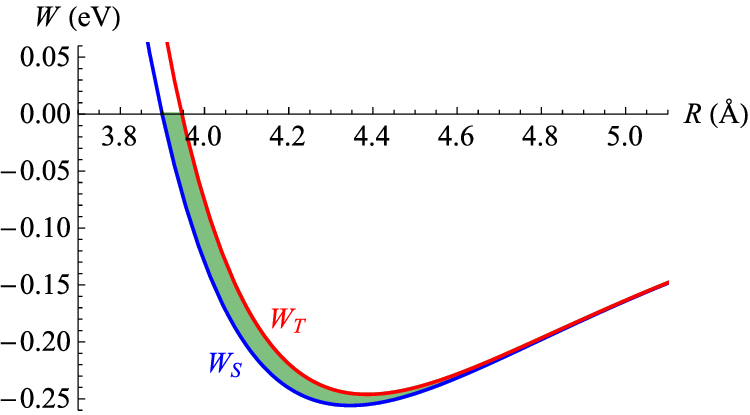}
 \caption{$W_S(R)$ and $W_T(R)$, eq. (\ref{WS-WT})
   [blue and red curves].
   The green area indicates the exchange interaction (\ref{g-def}).}
 \label{Fig-WS-WT}
\end{figure}

\noindent
The scattering length is given in Ref.\cite{vdW-Yb-PRA-08} 
in terms of a semiclassical (spin dependent) phase $\Phi$ 
by the following formula,
\begin{eqnarray}
  a_{\alpha} &=&
  \bar{a}~
  \Bigg\{
       1-
       \tan
       \bigg(
            \Phi_{\alpha}-
            \frac{\pi}{8}
       \bigg)
  \Bigg\},
  \label{a-Phi}
\end{eqnarray}
where $\alpha=S$ or $T$ for the two-atomic state with spin
wave function which is odd ($S$) or even ($T$) under
permutation of the atoms, and $\bar{a}$ is given by
eq. (\ref{bar-a}).
The semiclassical phase $\Phi_{\alpha}$ is defined
as \cite{vdW-Yb-PRA-08} [see eq. (\ref{Phi-def})],
\begin{eqnarray}
  \Phi_{\alpha} &=&
  \frac{\sqrt{M}}{\hbar}~
  \int\limits_{r_\alpha}^{\infty}
  \sqrt{-W_{\alpha}(R)}~
  dR,
  \label{Phi-append}
\end{eqnarray}
where
\begin{eqnarray}
  W_S(R) &=&
  W(R)-g(R),
  \nonumber \\
  W_T(R) &=&
  W(R)+g(R).
  \label{WS-WT}
\end{eqnarray}
$r_S$ or $r_T$ is a classical turning point for zero-energy
particle found from the equation $W_S(r_S)=0$ or
$W_T(r_T)=0$.

For a reference point, we also introduce the scattering
length $a_w$ for pure van der Waals potential (without
the exchange interaction). Of course, this quantity
cannot be experimentally measured,
\begin{eqnarray}
  a_{w} &=&
  \bar{a}~
  \Bigg\{
       1-
       \tan
       \bigg(
            \Phi_{w}-
            \frac{\pi}{8}
       \bigg)
  \Bigg\},
  \label{a-w}
\end{eqnarray}
where $\Phi_{w}(R)$ is given by eq. (\ref{Phi-w}).

\noindent
The potentials $W_S(R)$ and $W_T(R)$ are displayed in Fig.
\ref{Fig-WS-WT}. It is seen that $W_S(R)$ lies below
$W_T(R)$ which is evident due to the antiferromagnetic nature
of the exchange interaction. 


To proceed, we use $C_8$ and $\mu$ in Eqs. (\ref{vdW-def}) and (\ref{g-def})
as fitting parameters. For the best fit, we get
\begin{eqnarray}
  &&
  \Phi_S = 297.318,
  \ \ \ \ \
  \Phi_T = 291.521,
  \nonumber
  \\
  &&
  \Phi_w =  294.273,
  \label{Phi-STw-free}
\end{eqnarray}
and
\begin{eqnarray}
  &&
  a_S ~=~
  1878~a_B,
  \ \
  a_T ~=~
  219.7~a_B,
  \label{aS-aT-fit}
  \\
  &&
  a_w ~=~
  650.6~{\text{\AA}}.
  \label{aw-fit}
\end{eqnarray}
These values of $a_S$ and $a_T$ are close to the data
given by Eqs. (\ref{aS-experiment}) and (\ref{aT-experiment}).

\begin{figure}[htb]
\centering
\includegraphics[width=70 mm,angle=0]
   {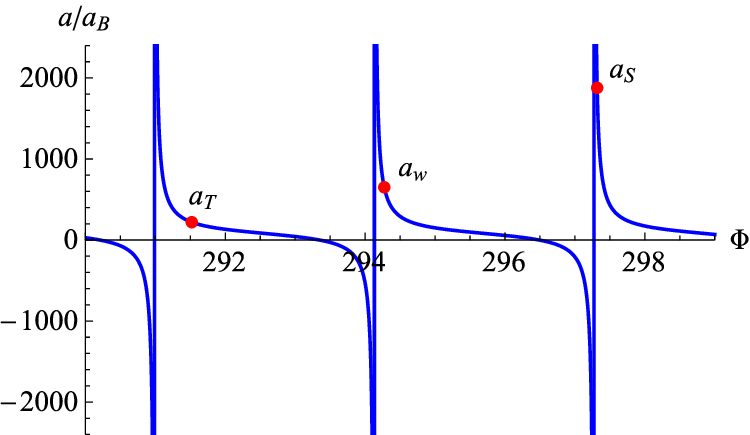}
 \caption{Scattering length as a function of the parameter $\Phi$
   (blue curve). The red points are $a_S$ and $a_T$,
   the scattering lengths (\ref{aS-aT-fit}) for the ``singlet'' 
   and ``triplet'' states, as well as $a_w$, the scattering
   length (\ref{aw-fit}) for the pure van der Waals potential. 
   The number of bound states increases by 1 as $\Phi$ increases 
   and crossed a singularity. According to the sketch in the figure 
   $N_S=N_T+1$. }
 \label{Fig-a-Phi}
\end{figure}

\subsection{Confinement-induced resonances}
  \label{subsec:CIR}

Following Ref. \cite{CIR-PhysRevA-2006}, we find ``singlet'' and ``triplet''
scattering lengths for itinerant Yb(\3P0) atoms scattered on trapped
Yb(\3P0) impurity. When scattering length for two free atoms is $a$,
scattering length for scattering of free atom on the impurity is
\begin{eqnarray}
  a_{\alpha} &=&
  2 a_{\alpha}^{(0)}
  \int \phi_0(r)
  \psi_{\mathrm{reg}}^{(\alpha)}(r)
  d^3 \mbfr,
  \label{scatt-length-eff}
\end{eqnarray}
where $\alpha = S, T$ for the ``singlet'' and ``triplet' two-atomic state and
$a_{\alpha}^{(0)}$ is the scattering lengths for two free atoms.
$\phi_{0}(r)$ is the wave function of the ground state of a 3D harmonic
oscillator describing the trapped atom, and
$\psi_{\mathrm{reg}}^{k = 0}(r)$ is defined as
\begin{eqnarray*}
  \psi_{\mathrm{reg}}^{(\alpha)}(r) &=&
  \lim_{\rho \to 0}
  \frac{\partial}{\partial \rho}
  \Bigg[
       \rho~
       \psi_{\alpha}
       \bigg(
            \mbfr + \frac{\boldsymbol\rho}{2},~
            \mbfr - \frac{\boldsymbol\rho}{2}
       \bigg)
  \Bigg].
\end{eqnarray*}
Here the two-atomic wave function $\psi(\mbfr_{\g},\mbfr_{\e})$ is found
from the equation,
\begin{eqnarray}
  \psi_{\alpha}(\mbfr_{\g},\mbfr_{\e}) &=&
  \phi_0(r_B) +
  g_{\alpha}
  \int
  G_{E}^{(\alpha)}(\mbfr_{\g},\mbfr_{\e};\mbfr,\mbfr)
  \nonumber \\ && \times
  \psi_{\mathrm{reg}}^{(\alpha)}(r)~
  d^3 \mbfr,
  \label{eq-LS}
\end{eqnarray}
where
$$
  g_{\alpha} =
  \frac{4 \pi \hbar^2 a_{\alpha}^{(0)}}{M}.
$$
Two-particle Green's function is
\begin{eqnarray}
  G_{E}^{(\alpha)}(\mbfr_{\g},\mbfr_{\e};\mbfr,\mbfr) =
  -\frac{M}{2 \pi \hbar^2}~
  \frac{1}{\big|\mbfr_{\g} - \mbfr\big|}
  \nonumber \\ \times
  \sum_{\mbfn}
  \phi_{\mbfn}(\mbfr_{\e})
  \phi_{\mbfn}(\mbfr)~
  e^{-\kappa_{\mbfn} |\mbfr_{\g} - \mbfr|},
  \label{Green-function}
\end{eqnarray}
where $\phi_{\mbfn}(\mbfr)$ are wave functions of 3D harmonic
oscillator in the state with harmonic numbers $\mbfn = (n_x,n_y,n_z)$,
$$
  \kappa_{\mbfn} =
  \sqrt{\frac{2 M \Omega_{\e}}{\hbar}~\big(n_x + n_y + n_z\big)},
$$
and $\Omega_{\e}$ is the harmonic frequency.

Before solving numerically Eq. (\ref{eq-LS}), we discuss,
why the confinement can give rise to resonances?
By the other words, why $a_{\mathrm{eff}}$ is singular?
Eq. (\ref{eq-LS}) can be formally written as
\begin{eqnarray}
  |\Psi_{\alpha}\rangle &=&
  |\Psi_{0}^{(\alpha)}\rangle +
  g_{\alpha} \hat{G}_{\epsilon} \hat{W} |\Psi_{\alpha}\rangle,
  \label{eq-LS-operators}
\end{eqnarray}
where the Green's function is $\hat{G}_{\epsilon} = (\epsilon - H_0 + i 0)^{-1}$, and
$$
  \hat{W}
  \Psi\bigg(\mbfR+\frac{\mbfr}{2},\mbfR-\frac{\mbfr}{2}\bigg) =
  \delta(\mbfr)~
  \frac{\partial}{\partial r}
  \bigg[
       r~
       \Psi\bigg(\mbfR+\frac{\mbfr}{2},\mbfR-\frac{\mbfr}{2}\bigg)
  \bigg].
$$
Formal solution of Eq. (\ref{eq-LS-operators}) is
\begin{eqnarray}
  |\Psi_{\alpha}\rangle &=&
  \frac{1}{1 - g_{\alpha} \hat{G}_{\epsilon} \hat{W}}~
  |\Psi_{0}^{(\alpha)}\rangle.
\end{eqnarray}

The operator $\hat{\mathcal{O}} = \hat{G}_{\epsilon} \hat{W}$ is
a Hermitian operator, and it has real eigenvalues $\lambda_i$,
$$
  \hat{\mathcal{O}} |e_i\rangle = \lambda_i |e_i\rangle,
$$
where $|e_i\rangle$ are eigenvactors of ${\mathcal{O}}$.
Then the scattering length (\ref{scatt-length-eff}) can be
written as \cite{CIR-PhysRevA-2006}
\begin{eqnarray}
  a_{\alpha} &=&
  2 a_{\alpha}^{(0)}~
  \bigg\langle
       \tilde\psi_{0}^{(\alpha)}
  \bigg|
       \frac{1}{1 - g_{\alpha} \hat{\mathcal{O}}}
  \bigg|
       \tilde\psi_{0}^{(\alpha)}
  \bigg\rangle
  \nonumber \\ &=&
  \frac{M}{4 \pi \hbar^2}
  \sum_{i}
  \frac{\big|\langle \tilde\psi_{0}^{(\alpha)} | e_i \rangle\big|^{2}}
       {g_{\alpha}^{-1} - \lambda_i},
  \label{scattering-length-vs-lambda}
\end{eqnarray}
where $\tilde\psi_{0}^{(\alpha)}(R) = R \Psi_{0}^{(\alpha)}(R,R)$.
Singularity in $a_{\alpha}$ is expected when $g_{\alpha}^{-1} = \lambda_i$,
or
\begin{eqnarray*}
  a_{\alpha}^{(0)} = \frac{M}{4 \pi \hbar^2 \lambda_i}.
\end{eqnarray*}
Since $\lambda_i$ is a function of $b_{\e}$ (the harmonic
length of the trapped atom), the last equation allow us to calculate
resonant values of $b_{\e}$.

\begin{figure}[htb]
\centering
\includegraphics[width=80 mm,angle=0]
   {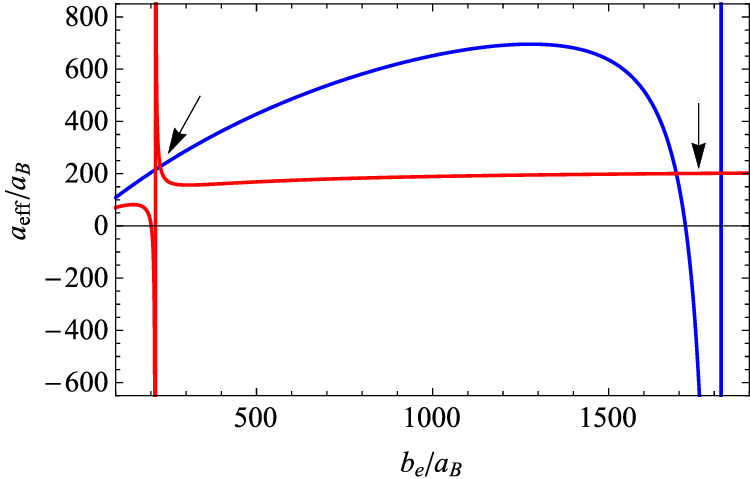}
 \caption{
   $a_{\mathrm{eff}}^{(S)}$ (blue curve) and
   $a_{\mathrm{eff}}^{(T)}$ (red curve)
   calculated numerically as functions of $b_{\e}$.
   The arrows show the intervals of $b_{\e}$ where
   $a_{\mathrm{eff}}^{(T)} > a_{\mathrm{eff}}^{(S)}$
   and the exchange interaction turns from
   ferromagnetic to antiferromagnetic.}
 \label{Fig-aSaT-eff}
\end{figure}

Solving Eq. (\ref{eq-LS}) numerically and substituting
$\psi_{\mathrm{reg}}^{(\alpha)}(r)$ into Eq. (\ref{scatt-length-eff}),
we get $a_{\alpha}$ for the ``singlet'' ($\alpha = S$) and
``triplet'' ($\alpha = T$) two-atomic states.
The results of the numerical calculations are shown in Fig. \ref{Fig-aSaT-eff}.
It is seen that for most part of the interval $b_{\e} < 1900 a_B$,
$a_{\mathrm{eff}}^{(S)} > a_{\mathrm{eff}}^{(T)}$, and the exchange
interaction is ferromagnetic.
However, there are two intervals, where
$a_{\mathrm{eff}}^{(S)} < a_{\mathrm{eff}}^{(T)}$, and the exchange
interaction is antiferromagnetic.
The first one is  close to a resonance for the ``singlet'' state,
$$
  1692 a_B < b_{\e} < 1819 a_B,
$$
and the second one is close to a resonance for the ``triplet'' state,
$$
  212.8 a_B < b_{\e} < 226.7 a_B.
$$
Corresponding harmonic frequencies are
\begin{eqnarray}
  &&
  0.3026~\mu{\text{K}} <
  \frac{\hbar \Omega_{\e}}{k_B} <
  0.3497~\mu{\text{K}},
  \label{singlet-res-num}
  \\
  &&
  19.48~\mu{\text{K}} <
  \frac{\hbar \Omega_{\e}}{k_B} <
  22.11~\mu{\text{K}}.
  \label{triplet-res-num}
\end{eqnarray}
Fig. 10 in Ref. \cite{CIR-PhysRevA-2006} show that
there are more resonances corresponding to higher
values of $\lambda_i$, but they are very narrow and
do not considered here.
In the following discussions, we demonstrate formation
of CIRs for $a_{\alpha}^{(0)} > 0$.

\subsubsection{Estimation of resonant values of $\Omega_{\e}$}

Here we follow Ref. \cite{CIR-PhysRevA-2006} and calculate
resonant frequencies for $a_{\alpha}^{(0)} \ll b_{\e}$.
In order to estimate resonant values of $\Omega_{\e}$,
we write two-atomic Hamiltonian as
\begin{eqnarray}
  H &=& H_R + H_r + H_{\mathrm{int}},
  \label{H=HR+Hr+HRr}
\end{eqnarray}
where $\mbfR = (\mbfr_{\g} + \mbfr_{\e})/2$ is the position
of the center of mass and $\mbfr = \mbfr_{\g} - \mbfr_{\e}$
is the relative coordinate.
\begin{eqnarray}
  H_R &=&
  -\frac{\hbar^2}{4 M}~
  \frac{\partial^2}{\partial \mbfR^2} +
  \frac{M \Omega_{\e}^{2}}{2}~R^2,
  \label{HR}
  \\
  H_r &=&
  -\frac{\hbar^2}{M}~
  \frac{\partial^2}{\partial \mbfr^2} +
  g_{\alpha}~\delta(\mbfr)~\frac{\partial}{\partial r}\big(r \cdot\big),
  \label{Hr}
  \\
  H_{\mathrm{int}} &=&
  \frac{M \Omega_{\e}^{2}}{8}~r^2 -
  \frac{M \Omega_{\e}^{2}}{2}~(\mbfR \cdot \mbfr).
  \label{H-int}
\end{eqnarray}
$H_r$ describe relative motion of the atoms with van der Waals
interaction (which is parametrized by the point-like interaction).
Consider the situation when the atoms are trapped by
the van der Waals interaction and form a molecule.
Energy level of the weakest-bounded $s$ state is
$$
  \epsilon_r =
  -\frac{\hbar^2}{M \big(a_{\alpha}^{(0)}\big)^{2}}.
$$
$H_R$ describes motion center of mass of the molecule
which is trapped by the optical lattice potential and possesses
harmonic oscillations.
Harmonic frequency of the molecule is
$\Omega_{\e}/\sqrt{2}$ and energies of the $s$ states are
$$
  \epsilon_R =
  \frac{\hbar \Omega_{\e}}{\sqrt{2}}~
  \bigg(2 n + \frac{3}{2}\bigg),
$$
where $n$ is an nonnegative integer.
Corrections to the energies from $H_{\mathrm{int}}$
[which are of order
$\hbar \Omega_{\e} \big(a_{\alpha}^{(0)}/b_{\e}\big)^{2}%
\ll \hbar \Omega_{\e}$]
are small and can be neglected hereafter.
Within this unperturbed approximation,
$a_{\alpha}$ diverges each time the energy of the oscillating
molecule is equal to the ground state energy
of the pair of atoms, i.e. at the values of
$\Omega_{\e} = \Omega_{n}^{(\alpha)}$ that satisfy
\begin{eqnarray}
  \bigg(
       2 n + \frac{3}{2}
  \bigg)~
  \frac{\hbar \Omega_{n}^{(\alpha)}}{\sqrt{2}} -
  \frac{\hbar^2}{M \big(a_{\alpha}^{(0)}\big)^{2}}
  =
  \frac{3}{2}~\hbar \Omega_{n}^{(\alpha)}.
  \label{eq-for-res}
\end{eqnarray}
Solving Eq. (\ref{eq-for-res}) we find resonant harmonic frequencies,
\begin{eqnarray}
  \hbar \Omega_{n}^{(\alpha)} &=&
  \frac{\hbar^2}{M \big(a_{\alpha}^{(0)}\big)^{2}}~
  \frac{4}{\sqrt{2}(4 n+3)-6)}.
  \label{Omega-res}
\end{eqnarray}
Taking into account Eqs. (\ref{aS-experiment}) and (\ref{aT-experiment}),
we get resonant frequencies for lowest $n$,
\begin{eqnarray*}
  \begin{array}{|c|c|c|}
    \hline
    n &
    \hbar \Omega_{n}^{(S)}/k_B (\mu{\text{K}}) &
    \hbar \Omega_{n}^{(T)}/k_B (\mu{\text{K}})
    \\
    \hline
    1 &
    0.2913 &
    21.29
    \\
    \hline
    2 &
    0.1189 &
    8.686
    \\
    \hline
    3 &
    0.0747 &
    5.456
    \\
    \hline
    4 &
    0.0544 &
    3.978
    \\
    \hline
  \end{array}
\end{eqnarray*}
The values for $n = 1$ are close to values (\ref{singlet-res-num})
and (\ref{triplet-res-num}) calculated numerically.


\end{document}